\documentclass[11pt,preprint]{aastex}
\newcommand{\temp}{$T$}
\newcommand{\hI}{H$^0$}
\newcommand{\hII}{H$^+$}
\newcommand{\heI}{He$^0$}
\newcommand{\neI}{Ne$^0$}
\newcommand{\cII}{C$^+$}
\newcommand{\cIIstar}{C$^{+*}$}
\newcommand{\oI}{O$^0$}
\newcommand{\nI}{N$^0$}
\newcommand{\mgI}{Mg$^0$}
\newcommand{\mgII}{Mg$^+$}
\newcommand{\arI}{Ar$^0$}
\newcommand{\kms}{km s$^{-1}$}
\newcommand{\logNHI}{$\log N(\mathrm{H}^0)$}
\newcommand{\NHtot}{$N(\mathrm{H}_\mathrm{tot})$}
\newcommand{\NHI}{$N($\ion{H}{1}$)$}
\newcommand{\NHII}{$N({\rm H^+})$}
\newcommand{\NHeI}{$N($\ion{He}{1}$)$}
\newcommand{\NOI}{$N($\ion{O}{1}$)$}
\newcommand{\NNI}{$N($\ion{N}{1}$)$}
\newcommand{\NCII}{$N($\ion{C}{2}$)$}
\newcommand{\NCIIstar}{$N($\ion{C}{2}$^*)$}
\newcommand{\NMgI}{$N($\ion{Mg}{1}$)$}
\newcommand{\NMgII}{$N($\ion{Mg}{2}$)$}
\newcommand{\NSiII}{$N($\ion{Si}{2}$)$}
\newcommand{\NSiIII}{$N($\ion{Si}{3}$)$}
\newcommand{\NSII}{$N($\ion{S}{2}$)$}
\newcommand{\NFeII}{$N($\ion{Fe}{2}$)$}
\newcommand{\NArI}{$N($\ion{Ar}{1}$)$}
\newcommand{\NArII}{$N($\ion{Ar}{2}$)$}
\newcommand{\HI}{\ion{H}{1}}
\newcommand{\HeI}{\ion{He}{1}}
\newcommand{\ArI}{\ion{Ar}{1}}
\newcommand{\nMgI}{$n({\rm Mg^0})$}
\newcommand{\nMgII}{$n({\rm Mg^+})$}
\newcommand{\nCIIstar}{$n({\rm C^{+*}})$}
\newcommand{\nCII}{$n({\rm C^+})$}
\newcommand{\nHI}{$n({\rm H^0})$}
\newcommand{\nHeI}{$n(\mathrm{He}^0)$}
\newcommand{\nOI}{$n(\mathrm{O}^0)$}
\newcommand{\nNI}{$n(\mathrm{N}^0)$}
\newcommand{\nNeI}{$n({\rm Ne^0})$}
\newcommand{\nel}{$n_{\rm e}$}
\newcommand{\scm}{cm$^{-2}$}
\newcommand{\cc}{cm$^{-3}$}
\newcommand{\epsC}{$\epsilon$ CMa}
\newcommand{\betC}{$\beta$ CMa}
\slugcomment{To be published in the Astrophysical Journal}

\begin{document}
\title{The Ionization of Nearby Interstellar Gas}

\author{Jonathan D. Slavin\altaffilmark{1}}
\affil{Eureka Scientific Inc., 2452 Delmer St. Suite 100, Oakland, CA
94602-3017}

\and

\author{Priscilla C. Frisch}
\affil{University of Chicago, Department of Astronomy and Astrophysics, 5460
S.\ Ellis Avenue, Chicago, IL 60637}

\altaffiltext{1}{Also: Harvard-Smithsonian Center for Astrophysics, 60 Garden
Street, MS 34, Cambridge, MA 02138}

\begin{abstract}
We present new calculations of the photoionization of interstellar matter
within $\sim 5$ pc of the Sun (which we refer to as the Complex of Local 
Interstellar Clouds or CLIC) by directly observed radiation sources including
nearby hot stars and the diffuse emission of the Soft X-ray Background (SXRB).
In addition, we model the important, unobserved EUV emission both from the hot
gas responsible for the SXRB and from a possible evaporative boundary between
the CLIC and the hot gas.  We carry out radiative transfer calculations and
show that these radiation sources can provide the ionization and heating of
the cloud required to match a variety of observations. The ionization
predicted in our models shows good agreement with pickup ion results,
interstellar absorption line data towards \epsC, and EUV opacity measurements
of nearby white dwarf stars.  Including the radiation from the conductive
boundary improves agreement with data on the temperature and electron density
in the cloud. The presence of dust in the cloud, or at least depleted
abundances, is necessary to maintain the heating/cooling balance and reach the
observed temperature.  Using the column density observations as inputs, we
derive the gas phase abundances of C, N, O, Mg, Si, S and Fe.  Based on these
inferred depletions, it appears that silicate and iron dust exists in the
CLIC, while carbonaceous dust has been destroyed. 
\end{abstract}


\section{Introduction}
The study of the interstellar medium (ISM) around and within the the Solar
System presents us with an unparalleled opportunity to gain understanding of
the equilibrium and abundances in warm interstellar gas.  Interstellar
neutrals, which are able to pass through the boundary between the solar wind
and the interstellar medium, constitute $\sim$98\% of the diffuse gas within
the heliosphere (the solar wind bubble), and form the parent population of the
pickup ions (PUI) and anomalous cosmic rays (ACR) seen within the heliosphere.
The PUI and ACR provide a direct sample of interstellar gas in the low density
cloud that surrounds the Solar System.  In turn, the interstellar gas flowing
past the Solar System determines the boundary conditions of the heliosphere.
By combining observations within the Solar System and towards nearby stars
with a theoretical understanding of the processes at work in the local ISM, we
can make progress in understanding both the warm interstellar medium and the
heliosphere.

The initial discovery of solar Ly$\alpha$ radiation fluorescing off of neutral
interstellar hydrogen within the solar system showed that the Sun is immersed
in warm tenuous neutral interstellar gas flowing through the solar system at
$\sim 25$ \kms \citep[although solar cycle variations in the radiation
pressure on the inflowing atoms produce uncertainties in fixing the cloud
velocity][]{ThomasKrassa:1971,BertBlam:1971,adfr77}.  Extreme ultraviolet
(EUV) observations of the backscattered \heI\ 584 \AA\ line provided
additional insight into the velocity and temperature of ISM within the solar
system, without radiation pressure uncertainties \citep{WellerMeier:1974}.
More recent data show that the source population of the pickup ions
\citep[e.g.,][]{Gloeckler:1996} and anomalous cosmic rays \citep[e.g.,][]
{Cummings:1996} is composed of interstellar neutrals that enter the
heliosphere, are ionized by either charge-exchange with the solar wind or
photoionization, and subsequently accelerated \citep[e.g.,][]{Fisk:1974}.  One
of the purposes of this study is to determine ionization levels for the
interstellar gas at the entry point of these neutrals into the heliosphere.

The overall characteristics of the nearest interstellar gas were determined by
the \emph{Copernicus} and IUE satellites.   Direct spectroscopic observations
by \emph{Copernicus} confirmed that the velocity of interstellar gas inside
and outside of the solar system is similar \citep{adfr77,Landsmanetal:1984}.
\emph{Copernicus}, IUE and HST observations of nearby stars show that average
space densities in local interstellar matter (ISM) are relatively low
\citep[$\sim 0.1$ \cc, e.g.,][]{McClintocketal:1975b,BruhweilerKondo:1982,
FrischYork:1983}, that the gas is relatively warm \citep[$\sim 7000$ K,
e.g.,][]{Landsmanetal:1984,Landsmanetal:1986b,Murthyetal:1987,Linsky:1996,
Lallement:1996} and that gas-phase abundances of refractory elements are
enhanced relative to cold cloud abundances indicating shock destruction of
dust grains \citep{fr81,Fea99}.  

Both optical and UV data show that even the nearest stars have multiple
foreground interstellar clouds, such as $\alpha$ Cen \citep[$d = 1.3$ pc, two
absorption components,][]{Landsmanetal:1984}, $\alpha$ Aql \citep[$d = 5$ pc,
three components,][]{Ferlet:1986}, and $\alpha$ CMa \citep[a.k.a.\ Sirius, $d
= 2.7$ pc, two components,][]{Lallementetal:1994}.  The velocity of the cloud
immediately surrounding the Solar System is established by \emph{Ulysses}
observations of interstellar \heI\ \citep[25 \kms, heliocentric
velocity,][]{Witte:1996}.  Absorption features at this velocity are identified
in many nearby stars \citep[e.g.,][]{Bertin:1993}, but puzzlingly not the
nearest star $\alpha$ Cen \citep{Landsmanetal:1984,LinskyWood:1996}.  Nearby
interstellar gas at the velocity of heliospheric interstellar \heI\ is
referred to as the Local Interstellar Cloud (LIC).  The two absorption
components present towards $\alpha$ CMa are at heliocentric velocities of 12
and 18 \kms\ \citep{Lallementetal:1994,Hebrard:1999}.  The 18 \kms\ component
corresponds to the LIC vector projected in this direction, and the
blue-shifted cloud (BC) velocity corresponds to a cloud at rest in the local
standard of rest \citep[for the LSR determined from \emph{Hipparcos} data of
13.4 \kms\ towards galactic coordinates $\ell = 28\degr$, $b = +32\degr$,][]
{DehnenBinney:1998}.  Given the extremely small distance to the BC component
and the relatively small velocity separation between the LIC and BC (less than
the LIC sound speed $\sim$10 \kms), we believe these clouds to be physically
adjacent. We refer to the complex of interstellar velocity components near the
Sun, including the LIC and BC components, as the Complex of Local Interstellar
Clouds \citep[CLIC, in place of the term Local Fluff used by][] {Frisch:1995}.
In this paper specifically, we denote the combined LIC and BC as the CLIC. 

The very local ISM offers clues to understanding the ISM in general since the
temperature and density of the CLIC are similar to that of the warm ionized
medium (WIM).  The WIM is a major constituent of the interstellar medium,
taking up $\gtrsim 20$\% of its volume and as much as 1/3 of its mass.  Much
of our knowledge of the state of the WIM comes from observations of diffuse
H$\alpha$ emission, diffuse emission from other optical lines including
[\ion{S}{2}] $\lambda$6717, [\ion{N}{2}] $\lambda$6584, [\ion{O}{3}]
$\lambda$5007, and [\ion{O}{1}] $\lambda$6300 \citep{HaffnerReynoldsTufte:1999}
and pulsar dispersion measures \citep[e.g., ][]{KulkarniHeiles:1987}.
These observations all involve integrations over long pathlengths and
therefore smooth out local variations in WIM properties.  The ionization of
the WIM inferred from such observations is considerably different from the
CLIC.  [\ion{O}{1}]$\lambda$6300 \AA\ observations have been used to infer that
the WIM (in the limited regions for which the line has been observed) is
highly ionized, $X_{\rm H} > 0.67$ \citep{R89}.  In addition, observations of
\ion{He}{1} $\lambda$5876 \AA\ have been used to infer that helium is
substantially \emph{less} ionized than hydrogen in the WIM, $X_{\rm He}
\lesssim 0.27 X_{\rm H}$ \citep{RT95}, though again the observations have been
limited to a few locations near the galactic plane.  

Pickup ion and absorption line data both indicate that the CLIC has only a
moderate ionization level (e.g.\ $X_\mathrm{H} \approx 0.3$) and observations
with \emph{EUVE} have shown He to be \emph{more} ionized than H \citep{Dea95}.
\HI\ 21 cm observations show widespread non-absorbing warm neutral gas with
temperatures generally in the range of 5000--8000 K
\citep[e.g.,][]{PayneSalpeterTerzian:1983,KulkarniHeiles:1988}.  This warm
neutral medium (WNM) is perhaps a better model for the CLIC, though it is
generally assumed that H is entirely neutral in such regions.
\citet{Heiles:2001}, however, has shown that $\sim 60$\% of \HI\ has $T>500$
K, with $\sim47$\% of this gas at temperatures considered thermally unstable
($T=500-5000$ K).  Low column density and low velocity warm clouds ($\log
N($\ion{H}{1}$) < 19$ \scm) are difficult to resolve where cold gas is
present, but are common in intermediate and high velocity halo gas
\citep[e.g.\ towards HD 93521,][] {SpitzerFitzpatrick:1993}.  

Generally, low emission measure ($<1$ cm$^{-6}$ pc) \ion{H}{2} gas occurs both
as ionized layers on neutral clouds \citep{Reynoldsetal:1995} and isolated low
column density ionized filaments \citep{Haffneretal:1998}.  The WIM may
represent a collection of regions with a range of ionization characteristics,
since  H-- and He--ionizing radiation is attenuated by an order of magnitude
for clouds with \logNHI$ \approx 17.6$ and $\approx 18.1$ \scm, respectively.
The hardness of the radiation field thus will vary considerably between clouds
with low neutral column densities (\logNHI$ < 18$ \scm).  Alternatively, the
CLIC may be characteristic of warm, ionized clouds in regions with little to
no O star radiation that are ionized primarily by radiation from hot gas.  

The ionization of nearby interstellar matter has been inferred from data on
\mgI/\mgII\ \citep{Bruhweileretal:1984,Frischetal:1990, Lallementetal:1994}
the fine-structure excited states of \cIIstar\ \citep{York:1983,
WoodLinsky:1997,Holberg:1999}, and observations of \HI\ and \HeI\ towards
nearby white dwarf stars \citep[e.g.,][]{Kimbleetal:1993,Dea95}.
\citet{Vallerga:1996} used \emph{EUVE} observations of \HI\ and \HeI\ column
densities combined with the directly measured value for \nHeI\ and showed that
both H and He are partially ionized with He most likely more ionized than H.
Electron densities have been found from \cIIstar/\cII\ and \mgI/\mgII\ ratios
but inconsistencies in the derived values ($\sim 0.11$ \cc vs.\ $\sim
0.1$--0.3 \cc) indicate the CLIC ionization has not been well understood up to
this point.  Recent studies of \arI\ abundances suggest that photoionization
dominates for nearby low column density gas \citep[based on arguments that
the large \arI\ photoionization cross-section reduces the \arI/\hI\ ratio below
ratios expected for recombining gas, ][]{SJ98,Jenkins:2000}.  

There is no single, clearly dominant source for the ionization of the CLIC.
The directly observed sources of hydrogen ionizing radiation fall into two
categories: stellar EUV sources and diffuse soft X-ray background emission
(SXRB). The former have all been observed by \emph{EUVE} and the combined
spectrum from the brightest sources has been presented by \citet{V98}. 
The most important part of the SXRB for ionization of the CLIC is the low
energy Be ($\sim 100$ eV) and B band ($\sim 175$ eV) radiation which has been
been observed by the Wisconsin Group using rocket-borne proportional counters.
\citet{V98} has shown that the stellar EUV sources are not capable of
providing the observed He ionization levels (27--50\%).  We show below that by
including the flux from the SXRB, modeled as emission from a $T \sim 10^6\,$K
collisional ionization equilibrium plasma, we \emph{can} account for the
observed ionization. We also show that better agreement with the observations
can be achieved if we include the radiation from an evaporative interface at
the boundary of the cloud.

\citet[][hereafter S89]{S89} explored ionization of the Local Cloud due to
ionizing radiation from the boundary of the cloud.  In S89 the detailed
temperature-density-ionization profiles at the interface of the cloud and the
hot gas of the Local Bubble were calculated assuming that conduction was more
or less inhibited by the magnetic field.  We have improved on the calculations
in S89 in several ways. First we treat the radiative transfer in the cloud
much more carefully, utilizing the photoionization code CLOUDY \citep{F96} for
this purpose.  In addition we use improved atomic data and codes in our
calculations of the radiation generated in the boundary and the resultant
ionization.  Since this study uses the ionization code CLOUDY to perform the
detailed radiative transfer calculation, our models do not make the usual
assumption that the cloud is uniformly ionized, allowing the ionization at the
solar location to be distinguished from values derived from absorption line
data toward nearby stars.  Moreover, substantial progress has been made in
determining the physical state of the cloud in recent years and the
differences in physical parameters from those assumed in S89 make a
substantial difference in the ionization calculations.

\section{Properties of the Nearby ISM \label{s-CLIC}}

\subsection{The Interstellar Ionizing Radiation Field}

The interstellar radiation field (ISRF) is the primary input to any model for
the ionization of low column density clouds such as the CLIC. The degree of
uncertainty in the intensity of the field varies greatly over the energy range
of importance for the ionization of the cloud ($\sim 7$--100 eV).  Although
still quite uncertain, perhaps the best determined part of the spectrum is the
far UV ($\sim 7$--13.6 eV) that comes primarily from B stars.  In our models we
have used the far ultraviolet (FUV) fields of \citet{MMP83} and \citet{GPW80}. 

The diffuse soft X-ray background (SXRB, 0.1--1 keV) is also very important to
cloud ionization, and has been observed over the entire sky by the Wisconsin
Group using sounding rockets \citep{Mea83} and with the \emph{ROSAT} satellite
\citep{Sea97}.  The limited energy resolution of the soft X-ray observations
does not allow tight constraints to be placed on the emission production
mechanism for the SXRB, although thermal emission from a hot plasma appears
most likely. Observations by different instruments have found consistent
results for the flux within the various energy bands.  Thus using the broad
band count rates to fix the intensity (i.e.\ the emission measure, see below)
in our model radiation field encourages confidence that interstellar
photoionization rates due to soft X-rays are fairly accurate.  Since the
radiation from an optically thin hot plasma is dominated by line emission,
however, we need to keep in mind that a coincidence of an emission line and an
absorption edge can still cause substantial differences in photoionization
rates calculated for spectra that produce the same band fluxes.  Of more
importance, however, is the extension of the hot gas emission spectrum down to
the EUV energies (i.e.\ 13.6--100 eV) where most of the ionization of elements
with first ionization potentials of 13.6 eV and higher occurs.  The chemically
abundant atoms of He (first ionization potential, FIP$ = 24.6$ eV) and Ne
(FIP$ = 21.6$ eV) which are found in the PUI population, and Ar (FIP$ = 15.84$
eV) which is observed both in the ACR population and in absorption in the ISM,
are sensitive to EUV fluxes.  

The EUV radiation field has important contributions both from the stellar flux
from white dwarfs and early type stars, and diffuse emission from plasma
interior to the Local Bubble.  Observations carried out with \emph{EUVE}
during the all-sky survey have determined flux from all of the brightest
stellar EUV sources \citep{V98}. The stellar flux (504--912 \AA) is dominated
by emission from two B stars, $\epsilon$ CMa and $\beta$ CMa.  The diffuse EUV
emission has been searched for but has not been clearly detected to date
\citep[see, e.g.][]{Jelinsky:1995,VS98}, though the observation is difficult
and no instrument optimized for observations of diffuse emission in the EUV
has yet been flown.  In this paper we take the simple approach of using the
SXRB observations to set the parameters of the Local Hot Bubble plasma model
(emission measure, $\int n_e n_{\mathrm{H}^+}\,ds$, and temperature) and then
use the emission spectrum predicted by the model at lower (EUV) energies.  We
discuss some of the uncertainties inherent in this approach below.  An
uncertain, yet possibly dominant, source of EUV emission is the radiation
generated within the interface boundary between the CLIC and Local Bubble
plasma as discussed below.

\subsection{\emph{In Situ} Observations of the ISM\label{insitu}}

The most direct observations of nearby interstellar material are observations
of interstellar gas and dust within the heliosphere, yielding cloud
temperature and space densities of abundant elements.  The direct detection of
\heI\ by \emph{Ulysses} yields $n($\ion{He}{1}$) = 0.017\pm0.002$ \cc\ and a
cloud temperature of $T = 7000 \pm 600$ K \citep{Witte:1996}.  The temperature
of \heI\ determined from observations of \ion{He}{1} 584 \AA\ backscattered
radiation, $T = 6900 \pm 600$ K, is consistent with this result \citep{Fea98}.
Although extensive data on interstellar \HI\ in the solar system exist, based
on Ly$\alpha$ backscattered radiation \citep[e.g.,][]{CBLK86}, somewhat
uncertain ionization and radiation pressure corrections and heliopause
filtration render these data less reliable for determining the properties of
the parent interstellar neutrals.  \emph{In situ} data on neutrals in the
solar system are listed in Table \ref{tab:obs}.

Ulysses observed He, Ne, N, and O pickup ions at distances $\sim$2--5 AU from
the Sun \citep{GloecklerGeiss:2001}.  These data are corrected for ionization
and propagation in the solar system, yielding the space densities of
interstellar \heI, \neI, \nI\ and \oI\ at the termination shock of the solar
wind.  These values are listed in Table \ref{tab:obs}.  However, up to 30\% of
interstellar \oI\ is lost during transition of the heliopause region by charge
exchange which couples interstellar \oI\ to the interstellar proton flow
deflected around the heliosphere by Lorentz forces
\citep{FahrRipken:1984,IML97}.  In pickup ions, \oI/\nI$ \sim 7.0$, compared
to the solar abundance ratio O/N$\sim$7.9 (Table \ref{tab:obs}).

The LIC contains interstellar dust, which has been directly observed by
detectors on both the \emph{Ulysses} and \emph{Galileo} satellites
\citep{Grunetal:1994}.  The dust grain size distribution and gas-to-dust mass
ratios ($R_{\rm gd}$) determined from these data are problematic for current
dust models, with \emph{in situ} data yielding $R_\mathrm{gd} < 140$ from
direct detections of interstellar dust grains, and $R_\mathrm{gd}\sim
425$--550 from missing-mass arguments applied to absorption line data
\citep[depending on assumed reference abundances,][]{Fea99}.  The upper limit
on $R_\mathrm{gd}$ determined from direct observations applies because small
dust grains are deflected around the heliosphere by Lorentz forces.  We find
below that the actual dust content of the cloud is not of great importance for
the heating/cooling balance since dust photoelectric heating is a minor
contributor ($\sim 2$\%) to the total heating rate.  (In contrast, the gas
phase abundances of the most abundant elements do have an important impact on
the cooling in the gas.)  The fraction of the abundant elements tied up in
dust provides important information on the nature of the dust in the LIC and,
by extension, in the WIM/WNM in general.

\subsection{Absorption Line Data\label{abun}}

Line of sight data, i.e.\ ion column densities, derived from observations of
nearby stars provide further constraints on our models of CLIC ionization.
HST high resolution (GHRS or STIS) or FUSE data exist for nearby A stars
\citep[e.g.\ $\alpha$ CMa,][]{Hebrard:1999}, several cool stars
\citep[e.g.,][]{Linskyetal:1995}, and white dwarf stars
\citep[e.g.,][]{Lemoineetal:1996,Holberg:1999,Jenkins:2000}.  A fairly
complete data set on the CLIC, is found for \epsC\ \citep[B2 Iab, $\ell, b =
239.8\degr$, $-11.3\degr$, $d = 132$ pc,][hereafter GJ]{GJ01}.  The stars
\epsC\ and $\alpha$ CMa (A1 V, $\ell, b = 227.2\degr$, $-8.9\degr$, $d = 2.6$
pc) are separated by $\sim13\degr$, and \epsC\ is located in the
third-quadrant void \citep[e.g.,][]{FrischYork:1983}.  The neutral gas
observed towards \epsC\ and $\alpha$ CMa shows the same velocity structure,
with the LIC component at $\sim 17$ \kms, and a second blue-shifted cloud (BC)
at $\sim 10$ \kms\ (heliocentric velocities).  These data indicate the LIC and
BC are both close to the Sun, and we choose to compare our model results with
observations of \epsC\ where data are more complete.  Additional ionized gas
towards \epsC\ appears to be more distant, contributing little to neutral
abundances (GJ).  \NHI\ is difficult to determine for low column density
sightlines because of saturation in the Ly-$\alpha$ line.  Values determined
for the total \NHI towards \epsC\ include \logNHI$ = 17.98\pm0.1$ \scm\
\citep[EUVE data,][]{Cassinelli:1995}, and \logNHI$ = 17.53$--17.91 \scm\ from
various abundance arguments applied to observed column densities (GJ).

The gas phase abundances of elements such as C, N, Mg and Fe in local gas can
not be determined directly from observations because while they are formed in
both neutral and ionized gas, there is no direct measure of \NHII\ in the
CLIC.  Previous studies of the CLIC show that it is neither nearly completely
ionized nor completely neutral, but is partially ionized with a significant
gradient in the ionization of the cloud from center to edge
\citep[e.g.,][]{ChengBruhweiler:1990,Vallerga:1996,SF98}.  Most importantly we
do not have any direct measurement of the degree of ionization of hydrogen,
although ionization equilibrium estimates using total line column densities
for \mgII, \mgI, \cII, and \cIIstar\ provide approximate values.  Because of
these uncertainties, our model treats the gas phase abundances of C, N, O, Mg,
Si, S and Fe as parameters which are varied to match observed column
densities.  Because of its proximity and the quality of the observational
data, we use column densities towards \epsC\ as determined by GJ for this
modeling. We include the combined LIC and BC components observed towards
\epsC, because both clouds contribute to the attenuation of the dominant
\epsC\ EUV radiation field.  Towards \epsC\, the LIC cloud has $\sim$54\% of
the neutral gas (GJ), while towards $\alpha$ CMa the LIC has $\sim$40\% of the
neutral gas \citep{Hebrard:1999}.  Line width constraints yield a BC
temperature of $\sim3000^{+2000}_{-1000}$ K, and LIC temperature
$\sim$8000$^{+500}_{-1000}$ K for $\alpha$ CMa.  Column densities towards
\epsC\ are given in Table \ref{tab:obs}.   In the modeling process, input
abundances are adjusted to force agreement with \NCIIstar, \NNI, \NOI, \NMgII,
\NSiII, \NSII\ and \NFeII.  The abundances of C, N, O, Mg, Si, S and Fe
therefore can be interpreted as a model result, and compared with our
expectations for elemental depletions and undepleted (``reference'')
abundances in the local ISM.

% added in response to referee's comments
We have selected the \epsC\ sightline for comparison because it is the lowest
column density star for which both \NCIIstar/\NCII\ and \NMgI/\NMgII\ data are
available.  Except for the presence of \ion{C}{4}, which has not yet been
% "observed" replaced with "detected" below 7/23/01
detected towards other nearby stars (and has had low upper limits placed on
its column density on a few lines of sight, see discussion in \S \ref{columns}
below), the \epsC\ sightline is representative of the nearby ISM.  For
instance, \NOI/\NNI$ = 8.3$ and 14.5 for the LIC and BC components,
respectively, compared to the range \NOI/\NNI$ = 4.0$--26.3 towards seven
nearby white dwarf stars and the two components towards \epsC\ and $\alpha$
CMa.  Nevertheless, as more data becomes available for other lines of sight,
it will be important to compare our models with those data sets.

\subsection{{Magnetic Field}}

In contrast to the increasingly tight constraints on the temperature and
density of the CLIC, the magnetic field strength in the cloud remains poorly
determined \citep{Frisch:1990,ratk98,Gloeckler:1997,Linde:2001}.  While
extremely high fields ($\ga 8 \mu$G) appear to be ruled out due to the lack of
detection of the heliospheric bow shock by \emph{Voyager 1}, the range of
plausible values for the field still extends from $\sim 1 - 5~\mu$G
\citep[e.g.,][]{Gayleyetal:1997,Linde:2001}.  An argument in favor of the
higher end of this range is that the pressure support provided by a field of
this strength would help support it against the apparently high thermal
pressure of the Local Hot Bubble.  Estimates of the bubble pressure based on
the observed soft X-ray emission put it at $P/k \sim 10^4$ \cc\ K
\citep{CR87,Bowyeretal:1995}. The thermal pressure in our cloud models is more
than a third less than that.  Nevertheless, the pressure determination for the
Local Bubble is indirect and subject to several uncertainties.  As a result we
explore models with $B = 2~\mu$G and $5~\mu$G to span the range of likely
values.  In one case (model 18), we use $3~\mu$G in an attempt to find a model
that better fit the variety of observational constraints.

\section{Photoionization Model}

\subsection{Radiation from an Evaporative Boundary}

If the Local Bubble gas is hot, $T \approx 10^6$ K, as inferred from the soft
X-ray background observations, then a sharp temperature gradient should exist
at the boundary between that hot gas and the warm, $T \approx 7000$ K LIC gas.
In such an interface thermal conduction will cause heat to flow into the cloud
and drive an evaporative outflow which produces mass loss from the cloud
\citep[see, e.g.,][]{CM77}.  (Maximum outflow speeds range from $\sim9$--36
\kms\ in our runs and depend strongly on the temperature of the hot gas and
weakly on the density of the cloud.) An important consequence is that the
cloud gas which is heated, ionized and accelerated outwards will radiate
strongly in the EUV.

Emission from the evaporative boundary is modeled using interface models
similar to those of S89. The code has been updated and improved, but all the
input physics is the same. We assume steady flow evaporation and spherical
symmetry and include the effects of radiative cooling, non-equilibrium
ionization and saturation of heat flux. The spectra (as well as necessary
ionization, recombination and cooling rates) are calculated using the Raymond
\& Smith plasma emission code \citep[][and updates]{RS77}.

The parameters that need to be specified for the evaporative interface models
include the cloud density, $n_{cl}$ (total density including H and He in all
ionization states), cloud radius, $R_{cl}$, temperature of the hot gas, $T_h$
(i.e.\ the temperature reached at some large $r$ -- we choose 30 pc -- from
the center of the cloud, effectively the temperature of the hot Local Bubble),
and cloud magnetic field strength, $B_0$.  In addition we specify a
conductivity reduction factor, $\eta$, which reduces the thermal conductivity
of the gas in the way that would occur if the mean field direction were at
some angle $\theta$ relative to the temperature gradient, where $\eta =
\cos^2(\theta)$.  We have chosen to always set $\eta = 0.5$ (corresponding to
$\theta = 45\degr$) which is the mean value for a field that could be at any
random angle to the radial direction. Note that this is different than
assuming a randomly tangled field which would result in sharply reduced
conductivity. The external ionizing radiation field also affects the cloud
evaporation rate due to the effects of ionization, particularly of H and He,
on the radiative cooling within the outflow.  In a steady flow, the heat
flowing into the cloud via thermal conduction is balanced by the radiative
cooling in the interface and the enthalpy flux out of the cloud.  Thus we also
need to specify parameters that influence the radiation field such as the
total \ion{H}{1} column density and the abundances of the most important
elements.

In all the cloud evaporation models presented below we have used $R_{cl} =
3\,$pc and $\eta = 0.5$.  The interface models have been run for values
$n_{cl} = 0.3$, 0.33 and 0.35; $B_0 = 2$ and $5\,\mu$G (except for model 18
which uses $B_0 = 3\,\mu$G) and $T_h = 10^6$ and $10^{6.1}\,$K.  In addition,
we have looked at the effects of varying the cloud column density, \NHI.  This
is allowed because \NHI\ is difficult to measure directly towards the B2 Ia
star \epsC\, and because \emph{EUVE} observations determine the total column
density towards the star while we are only interested in the contribution from
the CLIC.  That is, the \emph{EUVE} lines of sight could be, and in many cases
surely are, sampling clouds which do not attenuate the local radiation field.
We have evaluated models with \NHI$ = 4\times10^{17}$, $6.5\times10^{17}$ and
$9\times10^{17}\,$cm$^{-2}$.  \emph{EUVE} data gives total \NHI$ =
9.5\pm2.5\times10^{17}$ \scm\ towards \epsC\ \citep{Vallerga:1996}, while GJ
infer \NHI$ = 3.4-8.25\times10^{17}$ \scm\ from abundance arguments.  Our
choice of \NHI\ affects the degree to which the hot gas and interface emission
is absorbed between the edge of the CLIC and the Sun but does not affect the
stellar EUV emission received at the Sun.  This is because we start with the
observed flux and ``de-absorb'' by the amount appropriate to our assumed
column density for the CLIC to get the flux incident on the cloud face.  The
EUV flux incident at the cloud surface thus \emph{does} depend on the assumed
value for \NHI, with higher fluxes corresponding to larger \NHI\ values.
Twenty-five models have been created using these ranges of input variables,
and the parameter values for each individual run are listed in Table
\ref{tab:modparm}.  Included in these are seven models (nos.\ 19--25) in which
it is assumed that the cloud boundary is not conductive so that there is no
evaporation at the interface. In this way we are able to test the need for the
radiation from an evaporative interface in order to ionize the cloud.

\subsection{The Combined Radiation Field and Radiative Transfer}

To construct the total radiation field we take the cloud boundary spectrum we
have generated from the interface model and combine it with the de-absorbed
stellar EUV spectrum, the de-absorbed soft X-ray emission from hot gas and the
FUV field of either \citet{MMP83} or \citet{GPW80}.  The emission from the hot
gas is generated under the assumption of collisional ionization equilibrium of
an optically thin plasma using the \citet[][plus updates]{RS77} plasma
emission code. The total soft X-ray emission (including both the cloud
boundary emission and diffuse emission from the Local Bubble) is scaled so as
to give us the observed count rate in the Wisconsin B band ($\sim 175$ eV).
We choose to peg our flux to these observations since the B band is the
softest X-ray band for which there are observations that cover the entire sky.
We have also examined cases in which we assume no evaporation of the cloud so
that all the soft X-ray emission comes from the hot gas of the Local Bubble.
As we discuss below, our ionization models produce a worse match to the
observations for the no-interface cases.  We alter the diffuse part of the
flux somewhat to simulate, in a crude way, the radiative transfer effect of
having the diffuse emission coming from the entire sky.  This is done by
multiplying the flux (effectively the intensity integrated over the sky, $4
\pi J_\nu$) by $(1 + (1-\exp(-2\tau_\nu))/2\tau_\nu)/2$, where $\tau$ is the
optical depth (center to edge) of the cloud.  This expression is exact at the
edge of a spherical, constant optical depth cloud subject to a uniform diffuse
radiation field. 

\begin{figure}
\plotone{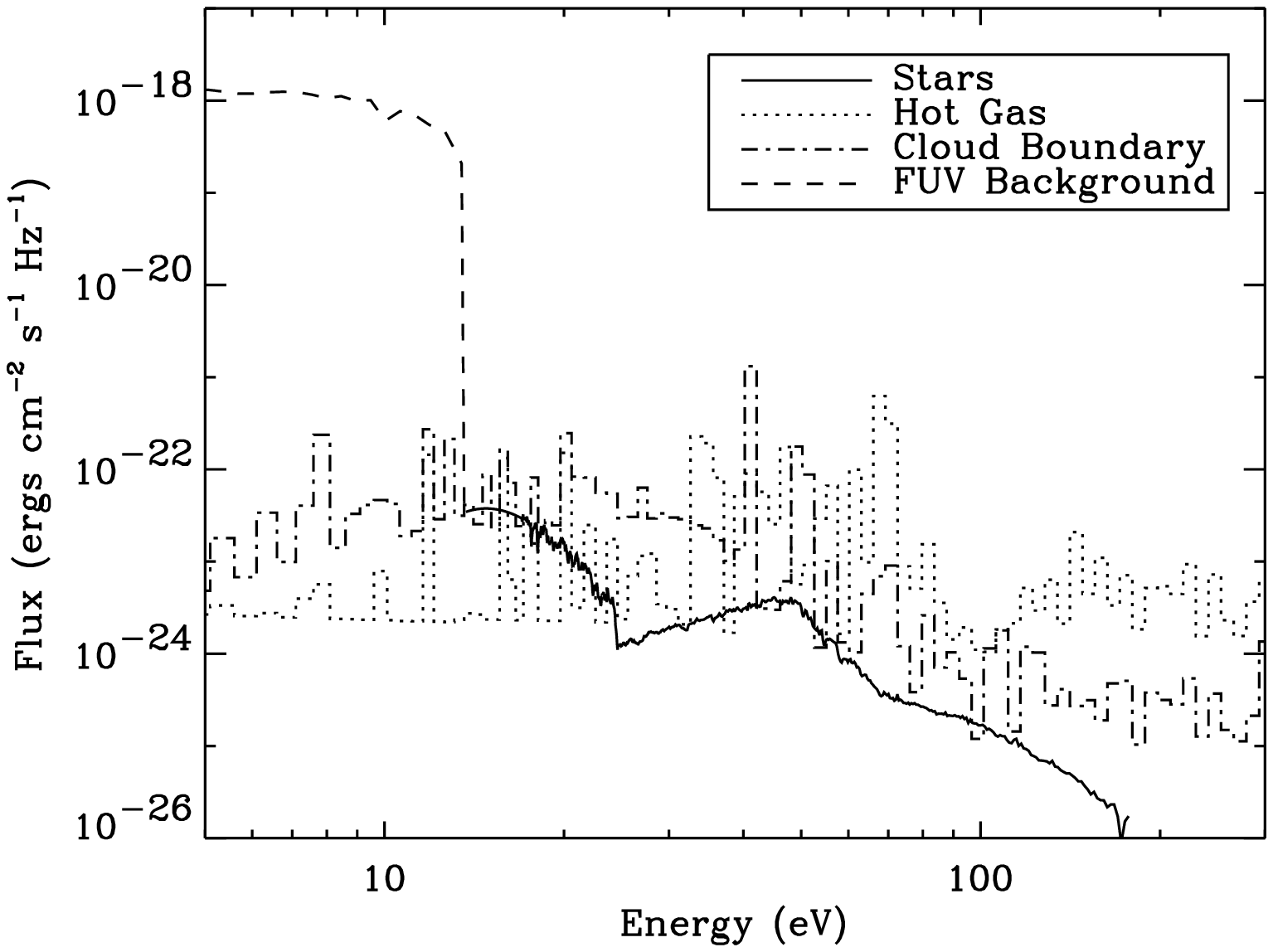}
\caption{Model for the interstellar radiation field incident on the complex of
local interstellar clouds (model no.\ 17).  The FUV part is mostly from B
stars \protect\citep{GPW80}.  The curve labeled ``Stars'' is the EUV flux from
nearby stars (WDs and B stars) observed by \emph{EUVE} \protect\citep{V98},
de-absorbed by an \protect\ion{H}{1} column density of $4\times10^{17}$
\protect\scm\ so as to get the flux incident from outside the cloud.  The
``Cloud Boundary'' curve is the flux from an evaporative interface between the
cloud and the hot gas of the Local Bubble.  The ``Hot Gas'' part of the
background is due to the $\log T = 6.1$ gas in the Local Bubble with the
intensity scaled so that the hot gas $+$ interface radiation is consistent
with the all-sky average count rate in the soft X-rays measured in the
Wisconsin B band \protect\citep{Mea83}.
\label{fig:radfield}}
\end{figure}

We show in Figure \ref{fig:radfield} an example of a radiation field
constructed for one of our models (no.\ 17, one of our preferred models
discussed below). In table \ref{tab:radfield} we list several measures of the
radiation field that characterize its hardness and intensity.  In the table,
$U$ is the ionization parameter, defined as the ratio of the ionizing photon
density to total (neutral$+$ionized) hydrogen density. The quantities
$\phi_\mathrm{H}$, $\phi_{\mathrm{He}^0}$, and $\phi_{\mathrm{He}^+}$, are the
total ionizing photon fluxes at the cloud surface in photons \scm\ s$^{-1}$ in
three different bands: 13.6--24.6 eV, 24.6--54.4 eV, and 54.4--100 eV
respectively.  $Q(\mathrm{H}^0)/Q(\mathrm{He}^0)$ is the ratio of the total
number of H and He$^0$ ionizing photons in the incident radiation field.
$\langle E\rangle$ is the mean energy (in eV) of an ionizing photon calculated
as $\langle E \rangle = \int F_\nu\; d\nu/\int \frac{F_\nu}{h\nu}\; d\nu$
where $F_\nu$ is the incident energy flux (ergs \scm\ s$^{-1}$ Hz$^{-1}$) and
the integration limits are from 13.6 eV to 100 eV.

% below added to address referee's comments
The spectrum of the emission from the hot gas and the cloud boundary is
dominated by lines and as a result, the best test of the model would be
observations of emission lines generated in the cloud boundary.  Up to now
no instrument has been had the sensitivity and spectral resolution in the EUV
to make such observations possible.  The upcoming CHIPS (Cosmic Hot
Interstellar Plasma Spectrometer) mission, however, will
have the capability to detect and resolve line emission from the cloud
boundary. In the CHIPS band (90\AA--260\AA) the primary emission lines from
the boundary are the \ion{He}{2} lines at 256\AA, 243\AA and 237\AA\ in order
of decreasing strength.  The strengths of these lines correlate strongly with
pressure in the hot gas (or total pressure, magnetic $+$ thermal in the cloud
gas) and weakly anti-correlate with \ion{H}{1} column density.  There is also
a significant, but non-linear correlation with temperature of the hot gas.
For the purposes of predicting the observations, the following rough
relationship holds for our models: $I($\ion{He}{2} 256\AA$) \approx P/k_B
0.006$ photons cm$^{-2}$ s$^{-1}$ sr$^{-1}$.  Here the pressure, $P/k_B$ in
cm$^-3$ K, is the thermal pressure in the hot gas or the total pressure in the
cloud. The 243\AA\ line is roughly a factor of 3 lower than the 256\AA\ line
and the 237\AA\ line is a factor of 3.8 down. Thus for $P/k_B = 10^4$
cm$^{-3}$ K, the 256\AA\ line is 60 photons cm$^{-2}$ s$^{-1}$ sr$^{-1}$, the
243\AA\ line is about 20 and the 237 line is about 16.  These intensities are
all high enough to be observable with CHIPS. Other lines in the CHIPS band
should give us insights into the nature of the bulk hot gas, but the helium
lines stand out as indicators of an evaporative cloud boundary.

We have employed the radiative transfer/thermal equilibrium code CLOUDY
\citep[version 90.05,][]{F96} in the mode in which plane-parallel geometry is
assumed.  CLOUDY calculates the detailed radiative transfer, including
absorption and scattering, of the incident field and the diffuse continuum and
emission lines generated within the cloud. Several options may be selected
when using the CLOUDY code. We have used commands to include cosmic ray
ionization (at the default background level), to assume constant pressure and
to include grains but at a 50\% abundance (command: grains 0.5). 
Temperature and ionization are calculated at each point within the cloud. We
have used the ionization at the end point, coinciding with the \NHI\ value
appropriate to our assumption for the generation of the radiation field, to
predict the cloud ionization at the heliosphere boundary.

\section{Model Results}

A range of assumptions were used in our modeling of the CLIC.  The parameters
of the models are listed in Table \ref{tab:modparm}, and model results
presented in Tables \ref{tab:cdres} and \ref{tab:ioniz}, and Figures
\ref{fig:neutdens} and \ref{fig:results2}.

\subsection{Column Densities and Abundances \label{columns}}

\begin{figure}
\plotone{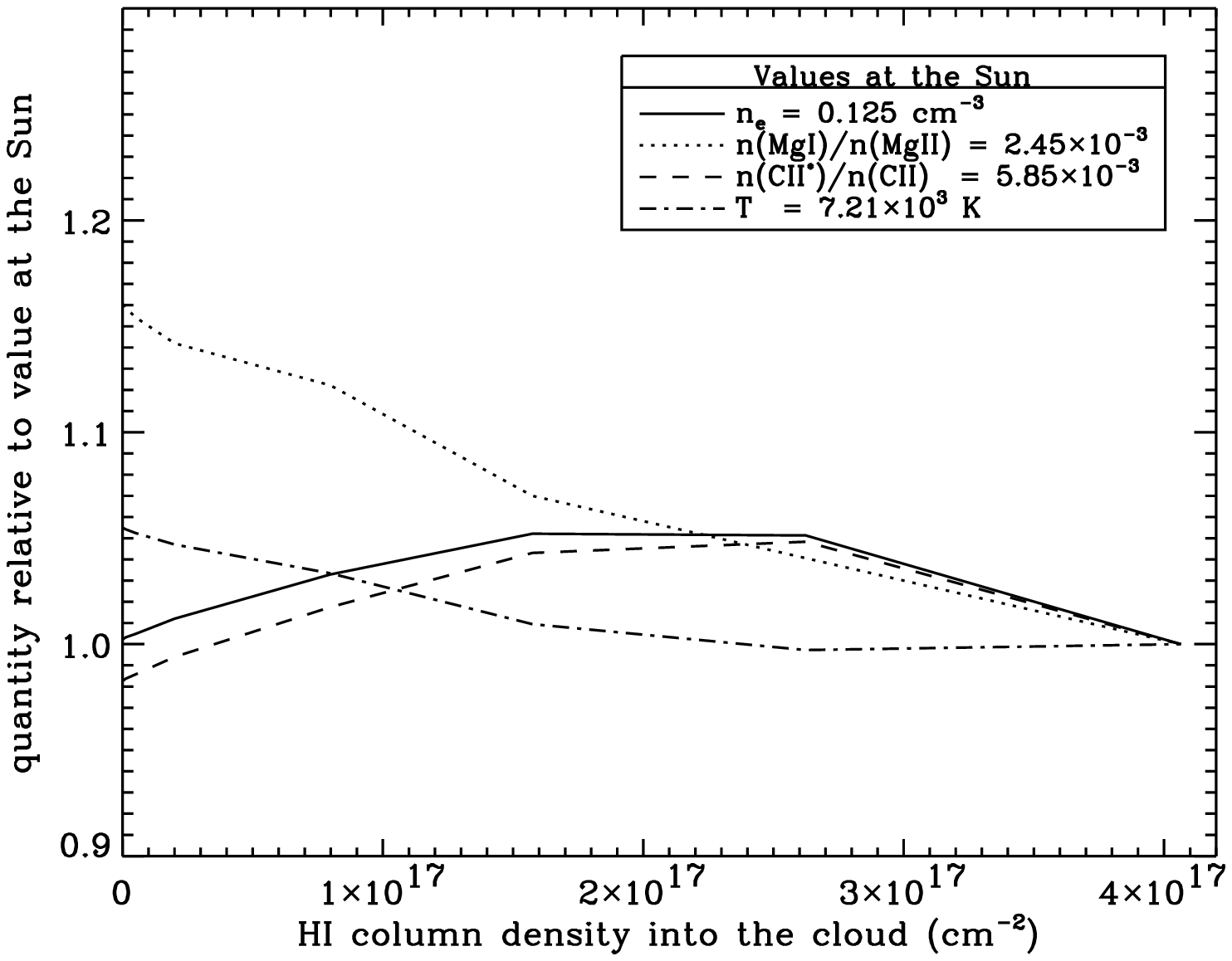}
\caption{Measures of the electron density in the cloud and temperature (for
our model no. 17) as a function of depth (\protect\ion{H}{1} column density)
into the cloud.  $n($\protect\ion{C}{2}$^*)/n($\protect\ion{C}{2}$)$ can be
seen to follow the true electron density closely while
$n($\protect\ion{Mg}{1}$)/n($\protect\ion{Mg}{2}$)$ shows substantial
deviation from \protect\nel. The increase in \protect\nel\ into the cloud is
due to a combination of increasing trapping of diffuse Lyman continuum flux
(leading to a slight increase in ionization) and an increase in total density
to counter the temperature decrease and maintain constant pressure (see
discussion in text).
\label{fig:ioniz}}
\end{figure}

In order to constrain our models and to derive information on the gas phase
elemental abundances in the CLIC we use HST observations of nearby
interstellar gas towards \epsC, which is similar to the ISM towards towards
$\alpha$ CMa (\S \ref{abun}).  As discussed above, we choose to use the
combined LIC and BC column densities for comparison.  The model was forced to
produce agreement with the observed column densities of \NCIIstar, \NNI, \NOI,
\NMgII, \NSiII, \NSII\ and \NFeII\ (Table \ref{tab:obs}) by adjusting input
elemental abundances to the model.  Model predictions were then compared with
\NMgII/\NMgI\ and \NCII/\NCIIstar, along with data from PUI and \heI\ within
the solar system (Table \ref{tab:obs}).  (Note: we list the Ulysses value for
$n($\ion{He}{1}$)$ \citep{Witte:1996} in Table \ref{tab:obs}, while PUI data
yield a somewhat lower value, $n($\ion{He}{1}$) = 0.0153\pm0.002$ \cc.
\citep{GloecklerGeiss:2001}) The abundances of \mgI\ and \cIIstar\ are
sensitive to electron densities, via radiative and dielectronic recombination
for \mgI, and collisional excitation of the fine-structure lines for \cIIstar.
The ratios \NMgII/\NMgI\ and \NCII/\NCIIstar\ provide good tests of our models
over spatial scales comparable to the cloud length.  Ratios of \NHI/\NHeI\
towards nearby white dwarf stars also provide good tests for the model
predictions, however large variations in this ratio are seen locally
\cite[e.g.][]{Dea95,Vallerga:1996,Wolffetal:1999}.  Table \ref{tab:cdres}
shows the predicted results for these ratios, as well as
$N(\mathrm{H_{tot}})$, \NArI, \NArII, and \NSiIII.  The variations of
\cIIstar/\cII, \mgI/\mgII\, \nel\ and $T$ with depth into the cloud are shown
for model no.\ 17 in Figure \ref{fig:ioniz}.

The ratio \NMgII/\NMgI\ depends on the electron density, yet it is also
sensitive to the strength of the FUV field since the ionization edge of \mgI\
is at $\sim 1622$ \AA.  Because the fraction of Mg that is neutral,
$X($Mg$^0)$, is always very small ($<$1\%), even minor (absolute) changes in
the FUV field can alter the ratio $X($Mg$^+)/X($Mg$^0)$ substantially.  For
this reason two different FUV fields have been explored, using the results of
\citet{MMP83} and \citet{GPW80}.  The FUV background of \citet{GPW80} was
based on direct observations of the radiation field with an extrapolation down
to 912 \AA\ done by theoretical calculations of stellar emission and dust
scattering and absorption. The intent of \citet{MMP83} was to describe the FUV
field in a more general way that would apply to the Galaxy at different
galactic radii.  The differences in the two model backgrounds are not great
(with the \citet{MMP83} flux somewhat larger than that of \citet{GPW80}), yet
they are large enough to affect the calculated \NMgII/\NMgI\ ratio in our
models.  As can be seen from comparison of the calculated values of
\NMgII/\NMgI\ in Table \ref{tab:cdres} with the observed values in Table
\ref{tab:obs}, in most cases the calculated ratio is larger than the observed
ratio.  This leads us to favor a lower FUV background flux, closer to that of
\citet{GPW80}.  Models 16 and 17 are identical except for the FUV radiation
field, which is MMP (16) and GPW (17).  The predicted ratios \NMgII/\NMgI\
differ by 13\% between the models, with Model 17 yielding better agreement
with data ($346\pm82$, Table \ref{tab:obs}).

In contrast to \cIIstar/\cII, the ratio \mgI/\mgII\ does not closely follow
the electron density.  This is clearly illustrated in Figure \ref{fig:ioniz},
which shows the predicted variations in \nel, \nMgI/\nMgII, \nCIIstar/\nCII,
and temperature with depth into the cloud.  (Note that the increase of \nel\
into the cloud is due to a combination of the increased trapping of Lyman
continuum recombination radiation and the increase in density in order to
maintain the constant pressure.) The difference in the spatial dependence of
\nel\ and \mgI/\mgII\ is due to the two things: the contribution of charge
exchange to the ionization rate of \mgI\ and the temperature dependence of the
recombination rate of \mgII. Charge exchange ionization of \mgI, i.e. via the
process \mgI\ + \hII\ $\rightarrow$ \mgII\ + \hI, is important for clouds like
the CLIC that are warm $T \sim 5000 - 10^4$ K with fairly high electron
densities, $n_e \approx 0.1$ \cc.  For our case the charge exchange ionization
rate coefficient is $\sim 10^{-10}$ cm$^3$ s$^{-1}$ at the cloud face as
compared with the photoionization rate of $4.7\times10^{-11}$ s$^{-1}$ for the
\citet{GPW80} FUV field. The sensitivity to temperature of the recombination
coefficent, mostly due to dielectronic recombination in this temperature
range, plays the biggest role in breaking the coupling between the electron
density and the \mgI/\mgII\ ratio.  For these reasons analyses that find the
electron density in clouds using \NMgI/\NMgII\ are actually finding an average
\nel\ weighted by the temperature dependence of the \mgII\ recombination
coefficient.  This is true to some degree even of sophisticated analyses such
as carried out by GJ who use both \NMgI/\NMgII\ and \NCIIstar/\NCII\ and
constrain both \nel\ and \temp.  An advantage of a radiative transfer
calculation such as we have carried out is that we do not need to assume a
constant \nel\ or \temp\ but can simply calculate the integrals directly.

A less ambiguous indicator for electron density is \NCIIstar/\NCII, which is
independent of the radiation field and weakly sensitive to temperature.  The
rate of excitation to the excited fine structure ($J = 3/2$) level of the
ground state of C$^+$ goes as $\sim T^{0.2}$ for $T \approx 7000$ K
\citep{BP92}.  Moreover cloud temperature is relatively well established (\S
\ref{insitu}).  \citet{Holberg:1999} find \nel$ =0.11^{+0.07}_{-0.06}$ for the
LIC from \cII\ fine-structure levels towards the white dwarf star REJ 1032+532
($\ell = 158\degr$, $b=+53\degr$, $d=132$ pc), while \citet{WoodLinsky:1997}
find \nel$=0.11^{+0.12}_{-0.06}$ towards Capella ($\ell=163\degr$,
$b=+5\degr$, $d=12$ pc).  These values tend to be smaller than typical
electron densities (0.11--0.5 \cc) found towards nearby stars (3--30 pc) from
\NMgII/\NMgI\ ratios alone \citep{Frisch:1994,Frisch:1995,Lallementetal:1994}.
GJ determine \nel$=0.12^{+0.05}_{-0.04}$ for the LIC, and \nel$=0.052\pm0.036$
for the BC, based on mutual constraints imposed by the \NMgII/\NMgI\ and
\NCII/\NCIIstar\ ratios.  The UV data also indicate the BC is evidently cooler
than the LIC (3000 vs.\ 8000 K) \citep{Hebrard:1999}.  We do not consider
these differences in LIC and BC properties here.  In future models we will use
a more accurate and detailed treatment of the cloud geometry.  

\begin{figure}
\plotone{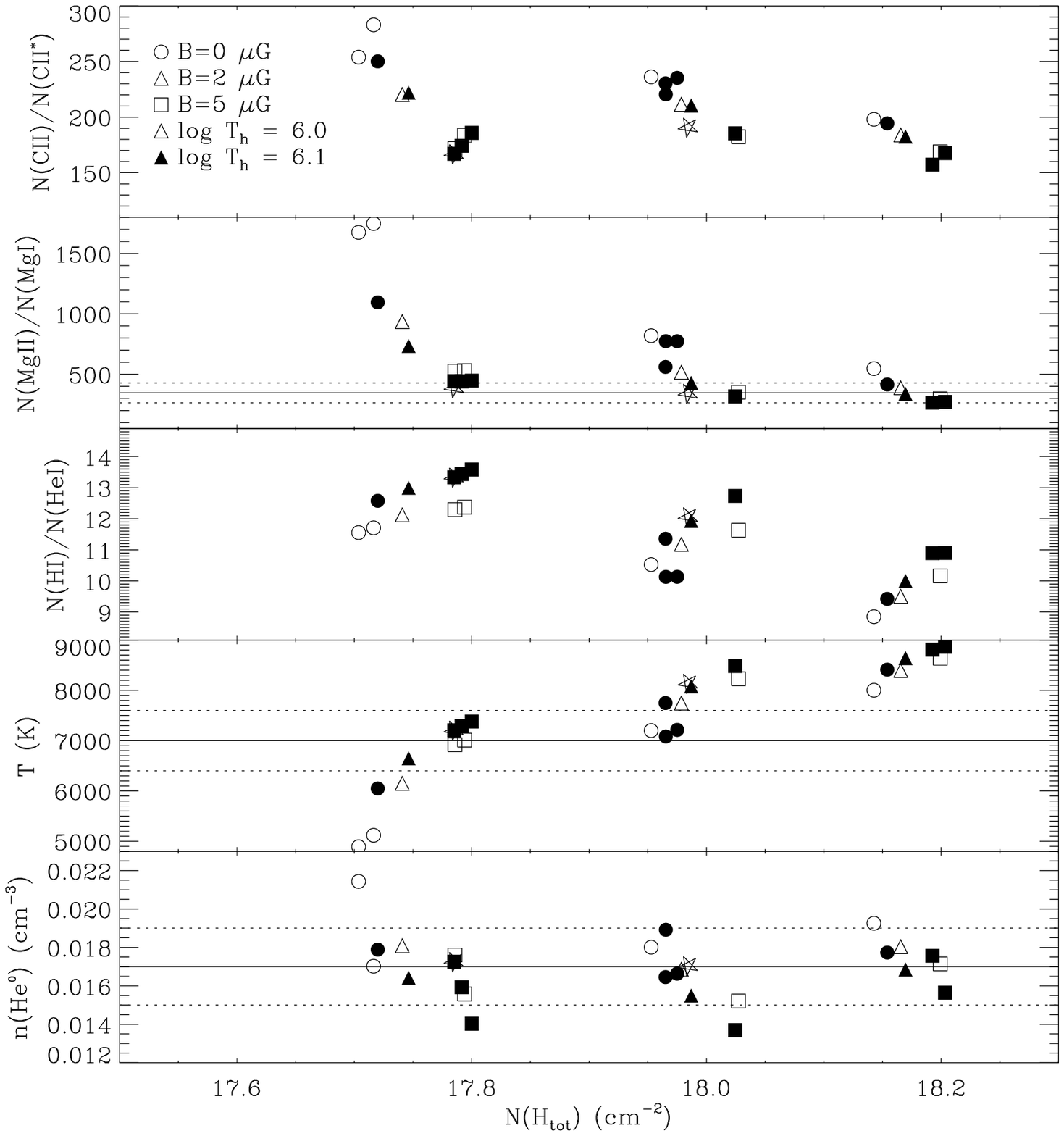}
\caption{Model results for column density ratios
\protect\NCII/\protect\NCIIstar, \protect\NMgII/\protect\NMgI, and
\protect\NHI/\protect\NHeI, and temperature and \protect\heI\ density at the
solar location plotted against the total column density
(\protect\hI+\protect\hII).  The three groups of $N($H$_\mathrm{tot})$
correspond to the three assumed values for \protect\logNHI$ = 17.60$, 17.81,
and 17.95 \protect\scm. The observed value (and limits) for
\protect\NMgII/\protect\NMgI\ towards \protect\epsC, $T$, and
$n(\mathrm{He}^0)$ are denoted by solid (value) and dashed (limits) lines. The
type of symbol denotes magnetic field strength in the model while
filled/unfilled denotes the value of $T_\mathrm{h}$.  The starred symbols are
for models with the \protect\citet{GPW80} FUV field (nos.\ 17 and 18) which
are also the models which fit the data best.  Note that cases in which the
same symbol appears more than once in a grouping differ only in the assumed
value of $n_\mathrm{H}$.
\label{fig:results}} 
\end{figure}

\begin{figure}
\plotone{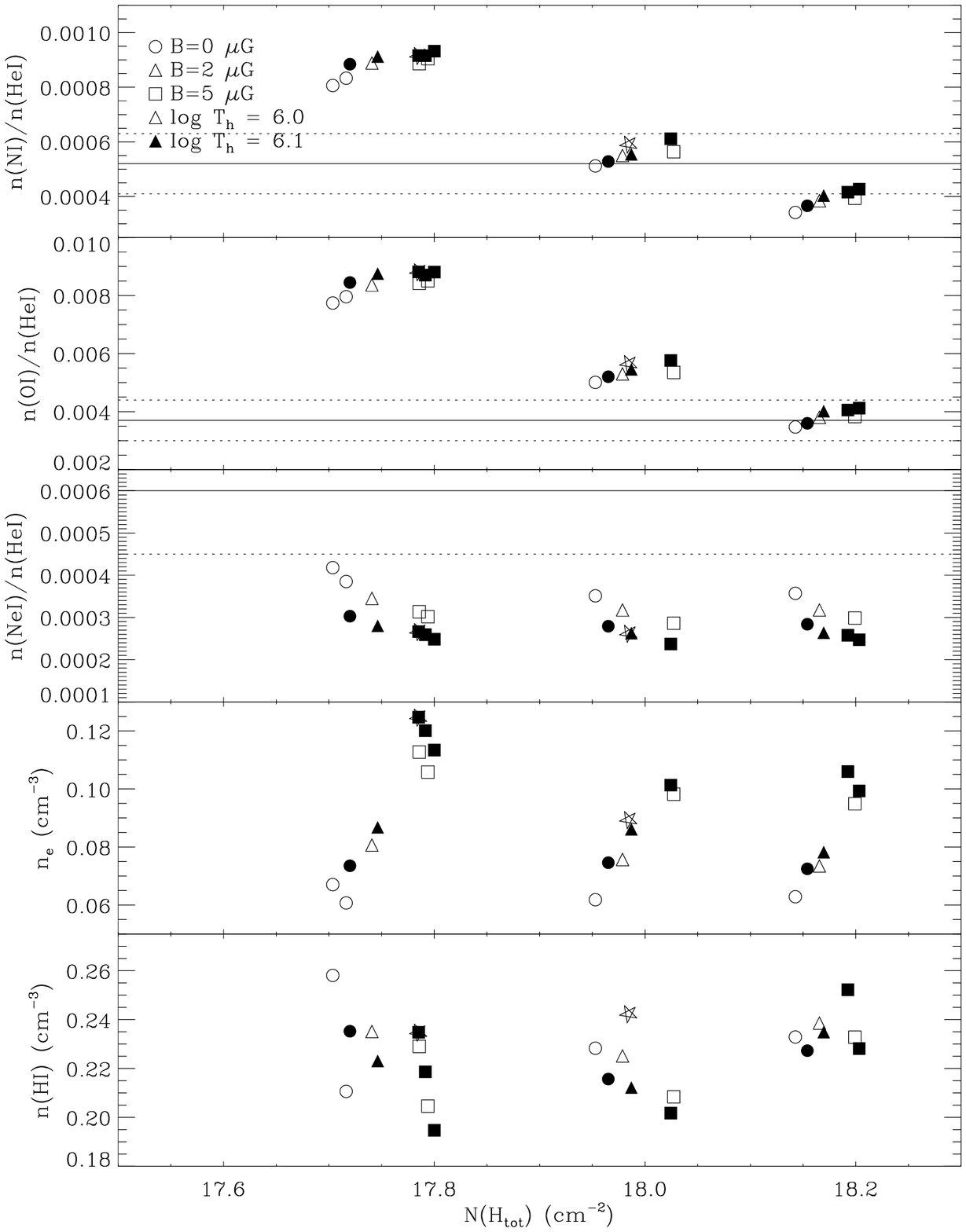}
\caption{Model results in the same format as Figure \protect\ref{fig:results}.
In this case we plot density ratios and densities for the solar location. The
observed values (and limits) from PUI data are shown plotted as solid lines
(dotted lines), from Table \protect\ref{tab:obs}.  The PUI data for O/He
should be corrected upwards by $\sim$20\% for comparison with model
predictions, to accomodate O filtration in the heliopause region (see text).
\label{fig:results2}}
\end{figure}

We find that the limits on \NCII/\NCIIstar\ (110--280) do not restrict the
models very well.  High electron densities are achieved for many of the
models, though interestingly not for models with no evaporative interface
(nos. 19--25).  As a result, models with no evaporative interface predict
\NMgII/\NMgI\ ratios that are factors of 2--3 larger than observed values.
The agreement with the observed \NMgII/\NMgI\ and \NCII/\NCIIstar\ ratios,
along with the predicted temperature and $n_{\mathrm{He}^0}$, cause us to
choose our model 17, with \nel$\sim 0.12$ at the Sun, as the ``best-fit''
model.  However, improvements in the uncertainties for column density data may
eventually allow better discrimination between the various evaporative
interface models.  Figure \ref{fig:results} and Figure \ref{fig:results2} show
the predictions of the models in graphical form.  Model no.\ 18 is in some
ways a better model than no.\ 17 (both are shown as stars in Figures
\ref{fig:results} and \ref{fig:results2}) but produces too high a temperature
at the solar location.  We discuss this further below.

The column density of \NArI\ is particularly interesting in that the
ionization of Ar is a good discriminant between photoionization equilibrium
models and non-equilibrium cooling models in which the local cloud shows the
signature of an earlier higher ionization state \citep{SJ98}. As \citet{SJ98}
show, if the Local Cloud had been highly ionized at some earlier epoch, e.g.\
by a strong shock, and is in the process of recombining, then \ion{Ar}{1} and
\ion{H}{1} will be roughly equally ionized since the recombination
coefficients for the ions are nearly the same.  The photoionization cross
section for \ion{Ar}{1} (15.8 eV), is substantially larger than for
\ion{H}{1}, so if photoionization is dominant, \ion{Ar}{1} should be
deficient.  \citet{Jenkins:2000} find a low \NArI/\NHI\ ratio for lines of
sight that include other gas in addition to the complex of local interstellar
clouds, favoring the photoionization equilibrium model over the fossil
ionization picture.  Our results show a range of values for \NArI/\NHI\ though
generally showing an even greater ionization of \ion{Ar}{1} ($\sim$75\%) than
inferred by \citet{Jenkins:2000}. Given the relatively large \HI\ column
densities, \NHI$=18.36$--18.93 \scm, observed towards these stars (G191-B2B,
GD~394, WD~2211-495, WD~2331-475), a substantial fraction of the \HI\ and
\ArI\ seen may not be associated with the CLIC.

Also of great interest is the column densities of highly ionized species that
are expected to exist in an evaporative cloud boundary.  Several studies,
motivated by UV observations of absorption lines for long lines of sight
through the disk and halo ISM, present calculations of the column densities
for \ion{C}{4}, \ion{N}{5}, \ion{O}{6} and \ion{Si}{4} in an evaporative
boundary \citep{CM77,BH87,BBF90}.  Our results for the models presented in the
previous tables are shown in Table \ref{tab:highions}.  These calculations
presume a line of sight that is radial in a spherical cloud and passes through
only one boundary.  Of the high ions listed above only \ion{C}{4} and
\ion{Si}{3} have been observed towards nearby stars. GJ derive a value for the
column density of \ion{C}{4} of $1.2\pm0.3\times10^{12}$ \scm.  Our model no.\
17 predicts $N($\ion{C}{4}$)\approx 2\times10^{12}$ \scm, which is about 60\%
larger than the GJ value.  Model no.\ 18 predicts a value of $N($\ion{C}{4}$)
= 1.3\times10^{12}$ \scm\ consistent with GJ's result.  The value of
$N($\ion{C}{4}$)$ depends on the level of conductivity in the interface
(parameter $\eta$ in the models, see above). Thus model 17 might be brought
into agreement with the observations by a small further reduction in $\eta$,
since a lower conductivity results in less mass loss.
% added to address concerns of the referee
If the magnetic field in the cloud is tangled, the conductivity could be
reduced by a large factor at all the boundaries of the cloud.  The degree to
which thermal conductivity is typically suppressed by magnetic field topology
in low density interstellar clouds is a difficult and unsolved problem.

We note that \citet{Bertin:1995} has put very low upper limits on the
\ion{C}{4} column density towards $\beta$ Leo, $\beta$ Car and $\beta$~CMa
($4.5\times10^{11}$, $3.2\times10^{11}$, and $4.9\times10^{11}$ \scm,
respectively).  Thus it would appear that the line of sight towards \epsC\ has
an exceptionally large \ion{C}{4} column density as compared to other nearby
lines of sight.  Further sensitive searches for \ion{C}{4} and \ion{Si}{4}
towards stars within the Local Bubble would help to constrain evaporating
cloud models.  As we discuss below, the observation of \ion{Si}{3} also is
problematic for evaporating cloud models.  Thus while we have shown that
evaporating cloud models do a good job of reproducing the ionization of the
cloud, it appears that the boundary surrounding the CLIC is more complex than
than a steady evaporative flow.

Table \ref{tab:highions} gives model predictions for $N$(\ion{Si}{3}), where
only the column density formed in the evaporative boundary is listed.  The
column density in the bulk of the cloud (Table \ref{tab:cdres}) turns out to
be of the same order of magnitude as in the boundary for our models. This is a
very interesting ion since it is not expected to be abundant in either
photoionized gas, which is too cold to produce Si$^{++}$ via collisional
ionization, or the cloud boundary gas in which Si is quickly collisionally
ionized up to Si$^{+3}$ and beyond. In regions like that in the CLIC in which
the radiation field is fairly diffuse, the ionization of Si is controlled by
the charge exchange reaction Si$^{++} +$ H$^{0} \rightleftharpoons$ Si$^{+} +$
H$^{+}$. Essentially all the Si is either Si$^+$ or Si$^{++}$ since Si$^0$ is
ionized by the FUV background. At $T \approx 7000$ K, charge exchange results
in a very small ionization fraction of Si$^{++}$, less than 1\% in our models.
GJ find column densities of $N($\ion{Si}{3}$) = 2.3 \pm 0.2 \times10^{12}$
\scm\ and $2.0 \pm 1.1\times10^{11}$ \scm\ for the LIC and BC components,
respectively.  This is far in excess of our calculated values
($4\times10^9$--$10^{10}$ \scm) for this ion.  This discrepancy would seem to
point to a need for a different type of interface than a classical evaporation
front.  One possiblity that deserves further exploration is that there is a
turbulent mixing layer \citep[see][]{SSB93} at the boundary between the CLIC
and the hot gas, at least in some parts of the interface.

The \NHI/\NHeI\ ratio observed towards nearby white dwarf star has been one of
the more difficult column density ratios to understand.  If stellar sources,
even very hot stars, are the dominant contributors to the interstellar
radiation field, one expects H to be substantially more ionized than He.  If
He has a 10\% abundance, however, the mean observed value of $\approx14$
\citep{Dea95,Frisch:1995} indicates that He is significantly
\emph{more} ionized than H in many sightlines.  Photoionization equilibrium
thus demands a rather hard spectrum, dominated by diffuse EUV and soft X-ray
emission with $E > 24.6$ eV.  Alternatively, more distant white dwarf stars
($>50$ pc) may sample gas with quite different radiation environments than the
CLIC, some of which may be in Stromgren spheres around hot white dwarfs
\citep{Wolffetal:1999,TatTerzian:1999}.  As can be seen from Table
\ref{tab:cdres}, our predicted values for \NHI/\NHeI\ range between 8.9 and
13.6.  Figure \ref{fig:results} shows that this variation is not a simple
function of total column density, although higher values of \NHtot\ tend to
have smaller \NHI/\NHeI\ ratios.  This is counter-intuitive, and occurs
because the EUV and SXRB fluxes are higher at the cloud surface for larger
values of \NHI\ since the de-absorption factor is greater.  Values approaching
the observed mean value of 14 result from models with higher temperatures for
the hot gas ($\log T_h = 6.1$) in accordance with the need for a relatively
hard background spectrum. This result is consistent with the results of
\citet{Sea98} who find the temperature of the emission from the Local Hot
Bubble to be $\log T = 6.07\pm 0.05$.

The gas phase elemental abundances of C, N, O, Mg, Si and Fe that we derive by
combining our model results with the observations are listed in Table
\ref{tab:abund}.  The elemental abundances ``derived'' in the model (found by
forcing predicted column densities to match observed column densities, \S
\ref{columns}) are listed in Table \ref{tab:abund}.  Since different models
have \HI\ column densities that differ by as much as a factor of 2.25
($4\times10^{17}$ \scm\ vs.\ $9\times10^{17}$), the abundances also vary by
more than a factor of 2 between different models.  Assuming a small value for
\NHI\ results in a high value for the abundances.  There are some general
features of our results, however.  O and C have close to the same abundance
with O up to 20\% higher for some cases. This is in contrast to their solar
abundance values \citep{G84} which put the O abundance at $\sim66$\% higher
than the C value. In addition, there are positive arguments that interstellar
abundances may be $\sim70$\% of solar abundances, and closer to B-star
abundances \citep{SnowWitt:1996,SavageSembach:1996}.  The predicted values for
C for the low \NHI\ cases exceed the standard solar abundance value of 490
parts-per-million (ppm), and are well above the standard ISM (i.e.\ gas phase)
value of 140 ppm \citep{Cea96}.  The B-star reference abundance for carbon is
only 240 ppm further exacerbating the problem.  Our results may point to the
need for a larger total (gas+dust) C abundance, at least for the very local
ISM.  This could be related to the larger ``carbon crisis'' of insufficient
carbon to explain both the dust and gas phase abundances of C in the ISM
\citep[see][]{KM96}.  The derived abundance for O (245--562 ppm) ranges from
31-70\% below the solar value indicating O atoms are depleted onto dust.  This
range encompasses the O abundance found for typical diffuse interstellar
clouds \citep[O/H$=319\pm14$ ppm][]{MeyerJuraCardelli:1998}.  The abundance of
N also slightly exceeds its solar value for the high \NHI\ cases.  Mg, Si and
Fe (solar abundances 38, 35 and 47 ppm respectively) on the other hand, are
substantially depleted in all cases.  Taken at face value, these results would
indicate that graphite or other carbonaceous grains have been destroyed in the
Local Cloud, but that silicate and iron grains have survived.

\subsection{Comparisons with Heliospheric ISM Data}

A novel aspect of this study is that we can compare model predictions with
\emph{in situ} observations of interstellar atoms within the solar system
(Table \ref{tab:obs}), including He, N, O, and Ne from pickup ion data
\citep[][hereafter GG]{GloecklerGeiss:2001}, and \nHeI\ and $T$ from Ulysses
data \citep{Witte:1996}.  Direct observations of $n$(\heI) determined from
\emph{in situ} observations of interstellar gas within the solar system
provide important constraints on cloud space density (Table \ref{tab:obs}).
Model predictions for the interstellar gas at the heliospheric boundaries are
presented in Tables \ref{tab:sunres} and \ref{tab:ioniz}, and Figures
\ref{fig:results} and \ref{fig:results2}.  The observed value, $n({\rm He^0})
= 0.017$ \cc, is higher than would be expected if the electron density were
low, $n_e \lesssim 0.1$ \scm\ consistent with our findings.  In addition, if
the \NHI/\NHeI\ column density ratio for the LIC+BC is close to the average
value of 14 \citep{Dea95}, then $n({\rm H^0}) \gtrsim 0.24$, since the
ionization of H will decrease into the cloud faster than that of He
\citep[e.g.,][]{Vallerga:1996}.  CLOUDY requires an initial input assumption
for the total H density (neutral$ + $ionized) at the outer edge of the cloud.
The models show that initial total space densities of $n \sim 0.3$ \cc\ or
more tend to produce a better match with observations of \heI\ in the solar
system.  

One interesting result is that despite large variations in model input
parameters, and differences in predicted abundances and other output
parameters, the ionization fraction of H at the position of the solar system
is found to be relatively constant between the models and within the range
$\sim 20$\% to 30\%.  The predicted He ionization ranges from $\sim 30$\% to
50\% at the solar location.  

The PUI data on O, N, Ne, and He \citep[][GG]{GG98} provide additional
constraints on cloud ionization.  We do not use ACR data
\cite[e.g.][]{Cummings:1996} for comparison \cite[e.g.][]{Frisch:1994},
because of the uncertain corrections for propagation and acceleration of ACR's
in the heliosphere.  Note that to derive the neutral fraction for elements
observed in the PUIs, one needs only to use data presented in Tables
\ref{tab:abund} and \ref{tab:sunres}. For element $Z$ (i.e.\ O, N or Ne) the
neutral fraction is $X(Z^0) = (n(Z^0)/n(\mathrm{He}^0))
(n(\mathrm{He}^0)/n(\mathrm{H}^0))(1-X(\mathrm{H}))/A_Z$ where $A_Z$ is the
abundance of $Z$. Absorption line data towards \epsC\ yield column density
ratios $N($\oI$)/N($\nI$) = 9.7^{+3.0}_{-1.9}$, while PUI data give space
density ratios of $n($\oI$)/n($\nI$) = 7.0\pm1.5$ at the termination shock
(Table \ref{tab:obs}).  Our model predictions for \nOI/\nNI\ (9.4--10.2 before
filtration) are close to the observed ratio from \epsC\ data (in part because
column densities were used to constrain the models).  However, up to 20--30\%
of interstellar \oI\ atoms may be ionized while crossing the heliopause region
by charge exchange ionization with interstellar protons compressed upstream of
the heliopause \cite[e.g.][]{FahrRipken:1984,Iz99}.  Some \nI\ ionization may
also occur in this region.  Both models 17 and 18 require at least $\sim$20\%
O filtration in the heliopause region to be consistent with the PUI data
(assuming N is unaffected by filtration). The constraint is actually looser
than this, however, if we take into account the fairly large uncertainty in
the \ion{O}{1} column density which was used to set the O abundance in the
models.

The variations of the model predictions for \nOI/\nHeI\ (0.004--0.009) are
larger than for the PUI O/He.  For model 17, which does very well fitting
other observational constraints, the \nOI/\nHeI\ ratio, 0.009, is
substantially above the PUI value, 0.0037.  In this case, an unrealistically
large filtration factor ($\sim 50$\%) is required to bring calculated and
observed values into agreement. Model 18, however, requires only about 20\%
O filtration.  Model 18 also agrees better with the N/He PUI ratio than model
17, with a predicted \nNI/\nHeI\ ratio ($5.9\times10^{-4}$) within the
uncertainties of the PUI value ($5.2\pm1.1\times 10^{-4}$).  Other models
predict values ranging from 4--$9\times10^{-4}$ for this ratio.

Finally, the models predict a value for \nNeI/\nHeI$ = 2.5-4.2\times10^{-4}$,
which is smaller than the observed value ($6.0\pm 1.5 \times10^{-4}$).  We
have assumed a Ne abundance of 123 ppm, based on solar abundances.  Based on
the results of these models, it appears that the Ne abundance in the ISM near
the Sun is significantly larger than the ``solar'' abundance of Ne (which is
based on meteoritic measurements).  There is some support for larger values
for Ne abundances, since 6 out of 7 planetary nebulae show Ne/O$ = 0.22 -
0.24$, versus the solar value of 0.16 \citep{Howardetal:1997}, suggesting a Ne
abundance of $\approx$180 ppm.  Based on the relative success of our best
models in matching both PUI and absorption line data, and the fact that the
ionization potentials of He (24.6 eV) and Ne (21.6 eV) are relatively close,
we conclude that the Ne abundance in the cloud around the solar system may be
$\sim 175$ ppm.

\begin{figure}
\plotone{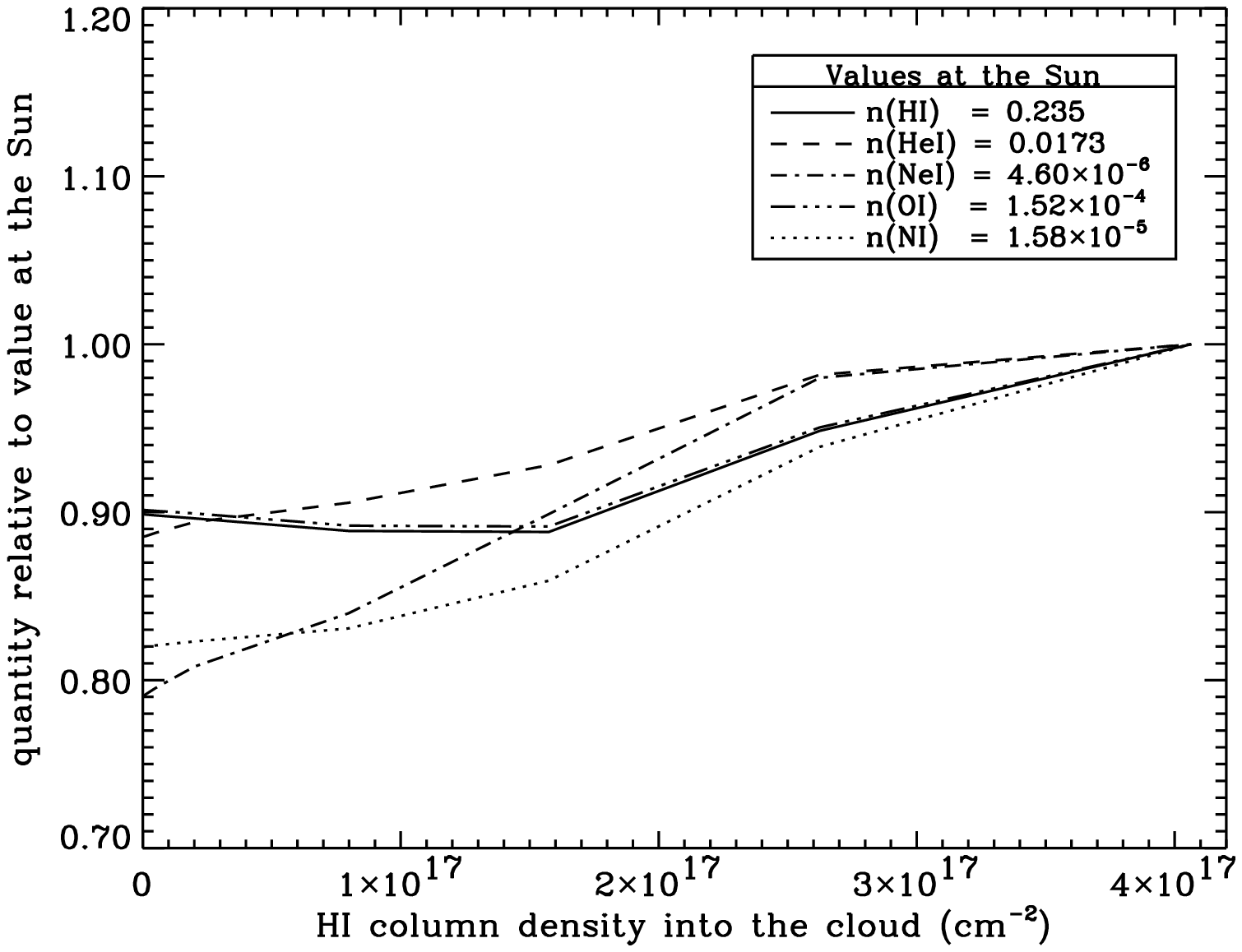}
\caption{Densities of \protect\ion{H}{1}, \protect\ion{He}{1},
\protect\ion{Ne}{1}, \protect\ion{O}{1}, and \protect\ion{N}{1} relative to
those at the Sun (for our model no. 17) as a function of depth
(\protect\ion{H}{1} column density) into the cloud.  The cloud surface is at
the left and the solar location is at the right.
\label{fig:neutdens}}
\end{figure}

In Figure \ref{fig:neutdens} we plot the density of \ion{H}{1},
\ion{He}{1}, \ion{Ne}{1}, \ion{O}{1} and \ion{N}{1} as a function of depth
into the cloud for model 17, one of our ``best fit'' models.  For almost all
of the neutral ions, density increases away from the cloud surface because the
ionization level decreases.  The degreee of variation of the densities
indicates how the position of the Sun within the cloud and the
column density of the cloud can affect neutral ion density ratios derived from
the PUI and ACR data. We list in Table \ref{tab:ioniz} the fraction of the
element in each ionization stage at the solar location (for model no.\ 17).
Figure \ref{fig:neutdens} demonstrates the tight coupling of O and H
ionization by charge exchange \citep{FieldSteigman:1971}.  The slight decline
in \nHI\ away from the cloud surface seen in the figure (and the increase in
\nel\ seen in Figure \ref{fig:ioniz}) is caused by the increased trapping of
Lyman continuum flux generated in the cloud by recombination in combination
with a decrease in temperature.  The increased trapping of Lyman continuum in
the cloud interior results in a lower effective recombination coefficient
which results in a flat or slightly increasing ionization fraction for H
despite the decreasing (due to absorption) photoionization rate.  The decrease
in temperature results in a higher total density (in order to maintain the
constant thermal pressure).  Deeper into the cloud, the effective
recombination rate approaches its case B value ($3.5\times10^{-13}$ cm$^3$
s$^{-1}$ at $T = 7000$ K) and the temperature flattens out leading to a
decrease in \nel\ and increase in \nHI\ near the cloud center.

The ionization of N is weakly coupled to H by charge exchange
\citep{ButlerDalgarno:1979}.  Both the charge exchange recombination
coefficient and the charge exchange ionization coefficient are roughly equal
to the radiative recombination coefficient ($\sim 10^{-12}$ cm$^3$ s$^{-1}$)
at the temperature of the LIC ($\sim 7000$ K).  The photoionization rate at
the cloud face for our model of the radiation field ($\sim 2\times 10^{-13}$)
is such that photoionization exceeds charge exchange ionization by a factor of
$\sim 2$.  Thus, while N ionization is influenced by H ionization, the
coupling is not very tight.  Photoionization and recombination of free
electrons alone would lead to a higher ionization fraction of N, while charge
transfer with H alone would lead to lower ionization fraction (roughly equal
to that of H).  Note that the important role of photoionization in the
ionization balance of N gives us an indication of the hardness of the
radiation field near $\sim 20$ eV since the photoionization cross section of N
is generally larger than that of H, rising from 2.4 times as large at
threshold to $\sim 19$ times at 100 eV.  

Observations of nearby white dwarf stars yield a range of values for
\NOI/\NNI, including $4.0 \pm 1.4$ towards REJ 1032+532  \citep[132 pc, 
\logNHI$ = 18.62$ \scm,]{Holberg:1999}, and 7--20 for four white dwarf
stars \citep[50--80 pc, \logNHI$ = 16.73 - 18.25$ \scm,]{Jenkins:2000}.  The
ratio \NOI/\NNI\ appears highly variable in low column density gas, and may
serve as a discriminant between mostly neutral versus partially ionized low
density (\NHI$\ll 10^{18}$ \scm) interstellar clouds.

The temperature of the Local Cloud turns out to be one of the more difficult
observations to match. Many models that appear acceptable in other ways
predict cloud temperatures that are substantially too high.  As we discuss
below, this could very well be due to errors and uncertainties in the atomic
data used by the codes to calculate our models.  Given the difficulties and
uncertainties present in thermal equilibrium calculations, we consider it
impressive that, without adjustment of parameters for this purpose, we are
able to come so close to matching the temperature of the cloud at the solar
system.  Figure \ref{fig:results} shows the temperature predictions of the
different models.

\section{Discussion}
\subsection{Model Assumptions and Reliability}
There are a number of assumptions made in our modeling of the Local Cloud that
may be questioned.  Perhaps foremost of these is the assumption of steady
state photoionization equilibrium \citep[see][]{LB96}.  The H recombination
time is $1/(\alpha^{(2)}n_e) \approx 9\times10^5$ yr, for $n_e = 0.1$ \cc\ and
$T \approx 7000$ K and it is quite likely that the Local Cloud has experienced
at least a moderately fast shock ($v_s \sim 50$ km s$^{-1}$) during that time.
The observations of a low \NArI/\NHI\ ratio by \citet{Jenkins:2000} referred
to above favor the interpretation that Ar is primarily photoionized and that
non-equilibrium recombination is not the dominant effect in determining the
ionization of the cloud.  We note in addition that our results show that the
local interstellar radiation field is quite capable of providing the
moderately high level of ionization that is observed for the CLIC.  Any fossil
ionization from an energetic event (e.g.\ the passage of supernova shock) in
the relatively recent past would appear to be insignificant at this point,
since the ionization of the cloud does not seem to be in excess of what we
expect from the ISRF.

One may also question the reliability and assumptions implicit in our
calculations of the radiation field from the hot gas and the evaporative
boundary.  Plasma emission codes are currently in a state of substantial
revision and new and more detailed atomic data are being incorporated into
these codes leading to significant changes in predicted spectra.  The Raymond
\& Smith code that we have used to generate the background radiation field is
known to be inaccurate in predicting a number of spectral features observed in
recent X-ray spectra using, e.g. ASCA and the Chandra X-ray Observatory. These
problems are of concern to us, though we feel, for the following reasons, that
for our purposes the inaccuracies in the code probably do not strongly affect
our results.  First, we are concerned only with the photoionization caused by
the background flux and not with the strengths of individual emission lines.
While individual spectral features could be incorrect, the cross section
averaged flux may be fairly accurate. Second, we scale the field strength to
be consistent with the observed band rate in the soft X-rays (i.e. the B
band), insuring that, at least over the range of the band coverage ($\sim 130
- 188$ eV), the photon flux is not far from the true value.  As more
observations of the diffuse background, particularly in the EUV, and updated
plasma emission codes become generally available, we will be able to revise
our background spectrum and reevaluate the ionization rate in the CLIC.

One particularly difficult aspect of the ionization calculation is the
treatment of the geometry of radiative transfer.  The sources of the
background radiation field include: stars, point sources distributed across
the sky (but dominated by \epsC\ and \betC); the hot gas of the Local Bubble,
roughly evenly distributed across the sky and generated from the volume of the
Bubble; and interface radiation generated in a thin volume between the warm
gas of the cloud and the hot gas of the bubble.  Each of these three sources
demands a somewhat different radiative transfer technique.  In addition the
Sun is not at the center of the complex of local interstellar clouds but
rather appears to be near the edge \citep{Frisch:1995}.  Moreover, the Blue
Cloud seems to have a lower temperature and somewhat higher density than the
LIC \citep{Lallementetal:1994,Hebrard:1999,GJ01}. Clearly the full radiative
transfer calculation in this situation would be extremely difficult and
subject to many uncertainties.  Nevertheless, a more complex geometrical model
may be warranted as more data on the shape and size of the Local Cloud and the
background radiation fields becomes available since these results show the
ionization at the solar system depends somewhat sensitively on the ionizing
flux recieved at our location within the cloud.  In future work we intend to
refine our treatment of the radiative transfer in the cloud and explore its
effects on the ionization at the solar location and throughout the cloud. In
addition we will explore ways to include the differences in LIC and BC
properties in our analysis.

\section{Conclusions}
We have presented results of a calculation of the ionization of the complex of
local interstellar clouds (LIC and BC) due to the background interstellar
radiation field.  The radiation field is constructed from directly observed
sources including nearby stellar sources (B stars and white dwarfs) and
diffuse emission from the hot gas in the Local Hot Bubble.  In addition,
emission from a proposed evaporative boundary between the warm cloud and
surrounding hot gas is included. Our results show that this radiation field is
capable of maintaining the ionization and heating necessary to explain a
variety of observations including: column densities of several ions towards
\epsC\, neutral atom ratios derived from PUI data, the temperature of the
cloud and the density of \ion{He}{1} observed in the solar system.

Our best models for the CLIC show good agreement with \NMgI/\NMgII,
\NCIIstar/\NCII\ ratios for the \epsC\ sightline.  The high mean electron
density inferred from the ratios \NCII/\NCIIstar\ and \NMgII/\NMgI\ towards
\epsC\ requires a local electron density at the Sun, $n_e \approx 0.1$ \cc.
This in turn requires a high EUV flux, larger than can be provided by either
the stellar EUV flux or the diffuse emission from the hot gas of the Local
Bubble.  Thus we find evidence for an evaporative boundary to the Local Cloud
as an additional source of EUV emission.  

One model, no.\ 17, fits the column density ratio data, and the \emph{in situ}
data on \temp\ and \nHeI.  The only significant problem for this model is the
ratio \nOI/\nHeI\, which is too large compared with the data.  Other potential
problems for that model include a large required C abundance (427 ppm) a large
magnetic field, 5 $\mu$G, and a small \HI\ column density, $4\times10^{17}$
\scm. Models 11 and 18 fit the column density ratio data, and the \emph{in
situ} \nHeI\ and the PUI data (with the exception of \nNeI/\nHeI, see
discussion below). Both models predict the temperature at the Sun to be
somewhat too high, however.  On the other hand, both models assume lower
magnetic fields and higher column densities and predict lower gas phase
abundances (e.g., 309 ppm for C for model 18) all of which are to some degree
more in keeping with our expectations.

By tying our results to observed column densities for a number of ions towards
\epsC, we are able to draw conclusions on the gas phase elemental abundances
of those elements Table \ref{tab:cdres}. We find that the gas phase abundances
of O, Mg, Si and Fe all show substantial depletion relative to solar
abundances.  N, S and especially C appear to be undepleted and even to have
abundances somewhat above the standard solar values.  Taken at face value we
would conclude that the LIC/BC complex has a significant amount of silicate
and, possibly Fe dust but that the carbonaceous dust has been destroyed.  The
results show clearly that for most elements both \hI\ and \hII\ must be
included when finding the intrinsic abundance of an element in low column
density clouds. 

We also find that our models, which are based on a gas phase abundance for Ne
of 123 ppm from solar abundances, predict substantially lower values for the
ratio \nNeI/\nHeI\ ($\sim3\times10^{-4}$) at the solar location than is
observed for the PUIs ($\sim6\times10^{-4}$).  Planetary nebulae data suggest
a Ne abundance of $\approx$175 ppm \citep{Howardetal:1997}, which is in better
agreement with our models.  This comparison suggests that ionization models
constrained by pickup ion data offer unique information on chemical abundances
in the ISM, although we can not claim yet that our models provide definitive
results.  However, the PUI data allow tighter constraints on the models than
available from absorption line data alone, and eventually will yield higher
standards of accuracy for these models.  Future data that would be very
helpful would be the determination of $n$(\arI)/$n$(\heI) at the entry point
to the heliosphere, either through new PUI data or through the interpretation
of existing Ar data in the anomalous cosmic ray population.

The relatively good agreement between predictions of our photoionization model
for local interstellar gas and both line-of-sight column densities towards
\epsC\ and \emph{in situ} observations of the ISM products in the solar system
strengthens our confidence that radiative transfer codes provide viable models
of clouds in space.  The radiation in the FUV, EUV and soft X-rays, play a
key role in maintaining ionization levels for a range of elements with a range
of ionization potentials.  Cloud interface emission (or another source with a
similar flux in the EUV) evidently must be included for accurate model
predictions that match data.  By restricting our models to the relatively well
known nearby interstellar gas, where we have the special situation that
\emph{in situ} observations of the space densities of selected elements are
available (including neutrals not seen in absorption), these models must be
considered better constrained than previous applications of ionization codes,
even though fundamental uncertainties remain.

The physical properties of the interplanetary environment within our solar
system, and in extra-solar planetary systems, are governed by the interaction
of stellar winds and the ISM \cite[e.g.][]{Frisch:1993}.  The heliospheric
configuration is sensitive to the pressure components of excluded charged
particles \citep{Holzer:1989}.  By combining an ionization model of the ISM
with observations of neutral interstellar He, N, O, Ne, and H atoms within the
heliosphere we are able to determine interstellar electron density immediately
outside of the heliosphere.  This photoionization model is especially useful
since the ionized component of the ISM can not be measured directly at the
solar location, yet provides a critical boundary condition of the heliosphere.

The interstellar cloud surrounding the solar system presents a unique
opportunity to determine the physics of a single low density warm interstellar
cloud, including both gas and dust components.  Low density clouds ($<
10^{18}$ \scm) such as the material surrounding the solar system generally are
seen only where high cloud velocities resolve weak individual components
\cite[e.g.][]{SpitzerFitzpatrick:1993,Wea99}.  It is essential that the
charged component of this cloud be understood before spacecraft are launched
to conduct \emph{in situ} studies of this ISM \cite[e.g.][]{ISP:2000}.  

\acknowledgments
This research was supported by NASA grants NAG 5-5077, NAG 5-6405,
and NASW-98027.  We have greatly benefitted from helpful discussions with Alan
Cummings, George Gloeckler, Dick Mewaldt, Dan Welty, and Gary Zank.  PCF would
also like to thank Carl Heiles and the Astronomy Department at the University
of California, Berkeley, for acting as a host during part of this research. We
thank the referee, John Vallerga, for comments that helped to improve this 
paper significantly.

%
% ---- Bibliography ----
%

\clearpage

\clearpage

\clearpage

\clearpage

\clearpage

\clearpage

%tab:obs
\begin{deluxetable}{llc}
\tabletypesize{\small}
\tablecolumns{3}
%\tablewidth{0pt}
\tablecaption{Observational Constraints \label{tab:obs}}
\tablehead{
\colhead{Observed} &  \colhead{Observed\tablenotemark{a}} &
\colhead{Reference} \\
\colhead{Quantity} &  \colhead{Value} & \colhead{} \\
}
\startdata
$N($\ion{C}{2}$)$ (cm$^{-2}$) & $2.1 - 3.4\times10^{14}$ & 1 \\
$N($\ion{C}{2}$^*)$ (cm$^{-2}$) & $1.5 \pm 0.31 \times10^{12}$ & 1 \\
$N($\ion{C}{4}$)$\tablenotemark{b} (cm$^{-2}$) & $1.2\pm 0.3\times10^{12}$ &
1 \\
$N($\ion{N}{1}$)$ (cm$^{-2}$) & $2.68 \pm 0.1 \times10^{13}$ & 1 \\
$N($\ion{O}{1}$)$ (cm$^{-2}$) & $2.6^{+0.8}_{-0.5} \times10^{14}$ & 1\\
$N($\ion{Mg}{1}$)$ (cm$^{-2}$) & $1.2 \pm 0.3 \times10^{10}$ & 1 \\
$N($\ion{Mg}{2}$)$ (cm$^{-2}$) & $4.15 \pm 0.11 \times10^{12}$ & 1 \\
$N($\ion{Si}{2}$)$ (cm$^{-2}$)  & $6.37\pm 0.3 \times10^{12}$ & 1 \\
$N($\ion{Si}{3}$)$ (cm$^{-2}$) & $2.5\pm0.31 \times10^{12}$ & 1 \\
$N($\ion{S}{2}$)$ (cm$^{-2}$) & $1.35 \pm 0.36 \times10^{13}$ & 1 \\
$N($\ion{Fe}{2}$)$ (cm$^{-2}$) & $1.87\pm 0.1 \times10^{12}$ & 1 \\
$N($\ion{H}{1}$)$/$N($\ion{He}{1}$)$ & $14\pm 0.4$\tablenotemark{c} & 2 \\
$n($\ion{O}{1}$)$/$n($\ion{N}{1}$)$\tablenotemark{d} & $7.0\pm 1.5$ & 3 \\
$n($\ion{O}{1}$)$/$n($\ion{He}{1}$)$\tablenotemark{d} & $3.7 \pm 0.7
\times10^{-3}$ & 3 \\
$n($\ion{N}{1}$)$/$n($\ion{He}{1}$)$\tablenotemark{d} & $5.2 \pm 1.1
\times10^{-4}$ & 3 \\
$n($\ion{Ne}{1}$)$/$n($\ion{He}{1}$)$\tablenotemark{d} & $6.0 \pm 1.5
\times10^{-4}$ & 3 \\
$T$(K)\tablenotemark{e} & $7000 \pm 600$ & 4 \\
$n($\ion{He}{1}$)$ (cm$^{-3}$) & $0.017\pm 0.002$ & 4 \\
\enddata
\tablenotetext{a}{\NCIIstar, \NNI, \NOI, \NMgII, \NSiII, \NSII\ and \NFeII\
are used to constrain the input abundances of the models.}  
\tablenotetext{b}{\ion{C}{4} is detected only in the LIC cloud.}
\tablenotetext{c}{The uncertainty given is only that due to uncertainties
listed in \citet{Dea95} for the observed \ion{H}{1} and \ion{He}{1} column
densities with the implicit assumption that the ratio is the same on all lines
of sight.  Given the substantial intrinsic variation in this ratio, however,
the quoted uncertainty must be regarded as a lower limit to the true
uncertainty.}
\tablenotetext{d}{Note that the pickup ion ratios are for values at the
termination shock.}
\tablenotetext{e}{The temperature shown represents the value derived from
in-ecliptic \ion{He}{1} data.  An alternative Ulysses sample from north of the
ecliptic yields a temperature of $6100 \pm 300$ K (Witte 1998, private
communication).}
\tablerefs{(1) \citet{GryDupin:1998,GJ01} (Note that the
values shown are for the LIC+BC combined towards \epsC.), (2) \citet{Dea95},
(3) \citet{GloecklerGeiss:2001}, Gloeckler, G., (2000, private communication),
(4) \citet{Witte:1996}.}
\end{deluxetable}

\clearpage
%tab:modparm
\begin{deluxetable}{cccccc} 
\tablecolumns{6} 
\tablewidth{0pc} 
\tablecaption{Model Input Parameter Values\label{tab:modparm}} 
\tablehead{ 
\colhead{} & \multicolumn{5}{c}{Input Parameter Type} \\ 
\cline{2-6} \\
\colhead{Model No.} & \colhead{$n_\mathrm{H}$ (cm$^{-3}$)} &
\colhead{$\log T_h$} & \colhead{$B_0$ ($\mu$G)\tablenotemark{a}} &
\colhead{ $N_\mathrm{HI}$ (10$^{17}$ cm$^{-2}$) } &
\colhead{FUV field\tablenotemark{b}} \\}
\startdata
1 & 0.273 &   6.0 &   5.0 &   4.0 & MMP \\
2 & 0.273 &   6.0 &   5.0 &   6.5 & MMP \\
3 & 0.273 &   6.0 &   5.0 &   9.0 & MMP \\
4 & 0.273 &   6.0 &   2.0 &   4.0 & MMP \\
5 & 0.273 &   6.0 &   2.0 &   6.5 & MMP \\
6 & 0.273 &   6.0 &   2.0 &   9.0 & MMP \\
7 & 0.273 &   6.1 &   5.0 &   4.0 & MMP \\
8 & 0.273 &   6.1 &   5.0 &   6.5 & MMP \\
9 & 0.273 &   6.1 &   5.0 &   9.0 & MMP \\
10 & 0.273 &   6.1 &   2.0 &   4.0 & MMP \\
11 & 0.273 &   6.1 &   2.0 &   6.5 & MMP \\
12 & 0.273 &   6.1 &   2.0 &   9.0 & MMP \\
13 & 0.300 &   6.0 &   5.0 &   4.0 & MMP \\
14 & 0.300 &   6.1 &   5.0 &   4.0 & MMP \\
15 & 0.300 &   6.1 &   5.0 &   9.0 & MMP \\
16 & 0.318 &   6.1 &   5.0 &   4.0 & MMP \\
17 & 0.318 &   6.1 &   5.0 &   4.0 & GPW \\
18 & 0.300 &   6.1 &   3.0 &   6.5 & GPW \\
19 & 0.273 &   6.0 & \nodata &   4.0 & MMP \\
20 & 0.273 &   6.0 & \nodata &   6.5 & MMP \\
21 & 0.273 &   6.0 & \nodata &   9.0 & MMP \\
22 & 0.273 &   6.1 & \nodata &   4.0 & MMP \\
23 & 0.273 &   6.1 & \nodata &   6.5 & MMP \\
24 & 0.273 &   6.1 & \nodata &   9.0 & MMP \\
25 & 0.227 &   6.0 & \nodata &   4.0 & MMP \\
\enddata
\tablenotetext{a}{Models for which no magnetic field strength is given are
those for which we have assumed that the cloud boundary is not conductive. For
these models there is no evaporative boundary.}
\tablenotetext{b}{Reference for FUV background field strength and shape. MMP
is \citet{MMP83} and GPW \citet{GPW80}.}
\end{deluxetable}

\clearpage
%tab:radfield
\begin{deluxetable}{ccccccc}
\tablecolumns{7}
\tablewidth{0pc}
\tablecaption{Characteristics of the Model Radiation Field\label{tab:radfield}}
\tablehead{
\colhead{Model} & \colhead{U} & \colhead{$\phi_\mathrm{H}$} &
\colhead{$\phi_{\mathrm{He}^0}$} & \colhead{$\phi_{\mathrm{He}^+}$} &
\colhead{$Q$(He$^0$)/$Q$(H$^0$)} & \colhead{$\langle E \rangle$}\\}
\startdata
1 & $3.9\times10^{-6}$ & $9.2\times10^{3}$ & $1.6\times10^{4}$ &
$5.6\times10^{3}$ & 0.49 & 57.8 \\
2 & $4.2\times10^{-6}$ & $1.2\times10^{4}$ & $1.5\times10^{4}$ &
$5.6\times10^{3}$ & 0.44 & 54.5 \\
3 & $5.5\times10^{-6}$ & $2.3\times10^{4}$ & $1.5\times10^{4}$ &
$5.6\times10^{3}$ & 0.33 & 44.8 \\
4 & $2.6\times10^{-6}$ & $5.3\times10^{3}$ & $9.0\times10^{3}$ &
$5.7\times10^{3}$ & 0.42 & 70.4 \\
5 & $3.0\times10^{-6}$ & $8.5\times10^{3}$ & $8.9\times10^{3}$ &
$5.7\times10^{3}$ & 0.36 & 63.3 \\
6 & $4.4\times10^{-6}$ & $2.0\times10^{4}$ & $9.0\times10^{3}$ &
$5.7\times10^{3}$ & 0.25 & 48.1 \\
7 & $4.4\times10^{-6}$ & $8.1\times10^{3}$ & $1.8\times10^{4}$ &
$8.4\times10^{3}$ & 0.50 & 59.9 \\
8 & $4.6\times10^{-6}$ & $1.1\times10^{4}$ & $1.7\times10^{4}$ &
$8.3\times10^{3}$ & 0.45 & 56.7 \\
9 & $6.0\times10^{-6}$ & $2.2\times10^{4}$ & $1.7\times10^{4}$ &
$8.3\times10^{3}$ & 0.34 & 47.3 \\
10 & $3.0\times10^{-6}$ & $5.1\times10^{3}$ & $1.0\times10^{4}$ &
$8.5\times10^{3}$ & 0.40 & 70.9 \\
11 & $3.4\times10^{-6}$ & $8.4\times10^{3}$ & $9.8\times10^{3}$ &
$8.5\times10^{3}$ & 0.35 & 64.7 \\
12 & $4.8\times10^{-6}$ & $2.0\times10^{4}$ & $9.9\times10^{3}$ &
$8.4\times10^{3}$ & 0.25 & 50.5 \\
13 & $3.6\times10^{-6}$ & $9.2\times10^{3}$ & $1.6\times10^{4}$ &
$5.5\times10^{3}$ & 0.50 & 57.7 \\
14 & $4.0\times10^{-6}$ & $8.2\times10^{3}$ & $1.8\times10^{4}$ &
$8.3\times10^{3}$ & 0.50 & 59.7 \\
15 & $5.5\times10^{-6}$ & $2.3\times10^{4}$ & $1.7\times10^{4}$ &
$8.2\times10^{3}$ & 0.35 & 47.2 \\
16 & $3.8\times10^{-6}$ & $9.8\times10^{3}$ & $1.8\times10^{4}$ &
$3.5\times10^{3}$ & 0.54 & 60.6 \\
17 & $3.8\times10^{-6}$ & $9.8\times10^{3}$ & $1.8\times10^{4}$ &
$3.5\times10^{3}$ & 0.54 & 60.6 \\
18 & $3.4\times10^{-6}$ & $1.0\times10^{4}$ & $1.2\times10^{4}$ &
$3.5\times10^{3}$ & 0.43 & 63.9 \\
19 & $1.9\times10^{-6}$ & $3.6\times10^{3}$ & $4.9\times10^{3}$ &
$5.8\times10^{3}$ & 0.32 & 83.4 \\
20 & $2.3\times10^{-6}$ & $7.0\times10^{3}$ & $4.8\times10^{3}$ &
$5.7\times10^{3}$ & 0.25 & 71.9 \\
21 & $3.6\times10^{-6}$ & $1.8\times10^{4}$ & $4.7\times10^{3}$ &
$5.7\times10^{3}$ & 0.16 & 50.8 \\
22 & $2.4\times10^{-6}$ & $3.7\times10^{3}$ & $6.3\times10^{3}$ &
$8.6\times10^{3}$ & 0.31 & 79.7 \\
23 & $2.8\times10^{-6}$ & $7.0\times10^{3}$ & $6.1\times10^{3}$ &
$8.6\times10^{3}$ & 0.26 & 71.0 \\
24 & $4.1\times10^{-6}$ & $1.8\times10^{4}$ & $6.0\times10^{3}$ &
$8.5\times10^{3}$ & 0.17 & 52.9 \\
25 & $2.3\times10^{-6}$ & $3.6\times10^{3}$ & $4.9\times10^{3}$ &
$5.8\times10^{3}$ & 0.32 & 83.4 \\
\enddata
\end{deluxetable}

\clearpage
%tab:cdres
\begin{deluxetable}{cccccccc}
\tabletypesize{\footnotesize}
\rotate
\tablecolumns{8}
\tablewidth{0pc} 
\tablecaption{Model Column Density Results\label{tab:cdres}} 
\tablehead{ 
\colhead{Model} & \colhead{$\log N(\mathrm{H_{tot}})$} & 
\colhead{$\log N($\ion{Ar}{1}$)$} & \colhead{$\log N($\ion{Ar}{2}$)$} &
\colhead{$\log N($\ion{Si}{3}$)$} &
\colhead{$\frac{N(\mathrm{Mg}\;\mathrm{II})}{N(\mathrm{Mg}\;\mathrm{I})}$} &
\colhead{$\frac{N(\mathrm{C}\;\mathrm{II})}{N(\mathrm{C}\;\mathrm{II}^*)}$} &
\colhead{$\frac{N(\mathrm{H}\;\mathrm{I})}{N(\mathrm{He}\;\mathrm{I})}$} \\}
\startdata
Obs.\tablenotemark{a} & \nodata & \nodata & \nodata & 12.40 &
$346\pm82$ & $110 - 280$ & $12 - 16$\\
\hline
1 & 17.79 & 11.44 & 11.96 & 10.20 & 529.3 & 183.7 & 12.38 \\
2 & 18.03 & 11.71 & 12.19 & 10.49 & 351.0 & 182.4 & 11.63 \\
3 & 18.20 & 11.93 & 12.38 & 10.63 & 297.3 & 168.8 & 10.16 \\
4 & 17.74 & 11.55 & 11.91 & \phn9.77 & 935.2 & 220.3 & 12.12 \\
5 & 17.98 & 11.81 & 12.15 & 10.22 & 515.8 & 211.5 & 11.17 \\
6 & 18.17 & 12.02 & 12.35 & 10.45 & 388.0 & 184.0 & \phn9.51 \\
7 & 17.80 & 11.42 & 11.94 & 10.31 & 447.4 & 185.9 & 13.59 \\
8 & 18.02 & 11.69 & 12.17 & 10.53 & 316.4 & 185.4 & 12.74 \\
9 & 18.20 & 11.91 & 12.36 & 10.68 & 270.9 & 167.8 & 10.90 \\
10 & 17.75 & 11.52 & 11.90 & \phn9.95 & 734.2 & 221.9 & 13.00 \\
11 & 17.99 & 11.78 & 12.14 & 10.33 & 428.7 & 210.5 & 11.93 \\
12 & 18.17 & 11.99 & 12.34 & 10.53 & 337.9 & 182.6 & 10.00 \\
13 & 17.79 & 11.46 & 11.95 & 10.15 & 525.8 & 171.8 & 12.29 \\
14 & 17.79 & 11.44 & 11.94 & 10.26 & 441.2 & 174.2 & 13.44 \\
15 & 18.19 & 11.92 & 12.36 & 10.64 & 264.3 & 157.2 & 10.90 \\
16 & 17.79 & 11.45 & 11.93 & 10.23 & 441.9 & 167.0 & 13.34 \\
17 & 17.79 & 11.45 & 11.93 & 10.23 & 389.2 & 167.0 & 13.34 \\
18 & 17.98 & 11.77 & 12.14 & 10.34 & 341.9 & 190.9 & 12.06 \\
19 & 17.70 & 11.64 & 11.87 & \phn9.08 & 1675\phd\phn\phn & 253.9 & 11.56 \\
20 & 17.95 & 11.89 & 12.12 & \phn9.92 & 819.3 & 236.2 & 10.53 \\
21 & 18.14 & 12.10 & 12.32 & 10.22 & 546.9 & 198.0 & \phn8.85 \\
22 & 17.72 & 11.58 & 11.87 & \phn9.64 & 1095\phd\phn\phn & 250.1 & 12.58 \\
23 & 17.97 & 11.85 & 12.12 & 10.16 & 560.6 & 230.3 & 11.36 \\
24 & 18.15 & 12.06 & 12.33 & 10.39 & 415.3 & 194.4 & \phn9.42 \\
25 & 17.72 & 11.61 & 11.88 & \phn9.24 & 1748\phd\phn\phn & 283.0 & 11.71 \\
\enddata
\tablenotetext{a}{Observational results from \citet{GJ01} (see table
\ref{tab:obs}). The values listed for \NHI/\NHeI\ are the range of values
observed excluding Feige 24 which is one of the most distant stars observed by
\citet{Dea95} and has unusually large \NHI\ and ratio values.}
\end{deluxetable}

\clearpage
%tab:highions
\begin{deluxetable}{cccccc} 
\tablecolumns{6}
\tablewidth{0pc}
\tablecaption{High Ion Column Densities For Models Including Cloud Evaporation
\tablenotemark{a}\label{tab:highions}}
\tablehead{
\colhead{Model} & \colhead{$\log N($\ion{C}{4}$)$} & 
\colhead{$\log N($\ion{N}{5}$)$} & \colhead{$\log N($\ion{O}{6}$)$} &
\colhead{$\log N($\ion{Si}{4}$)$} & \colhead{$\log N($\ion{Si}{3}$)$} \\}
\startdata
1 & 12.35 & 11.89 & 13.03 & 10.27 & 10.12 \\
2 & 12.11 & 11.68 & 12.81 & 10.05 & \phn9.90 \\
3 & 11.92 & 11.53 & 12.68 & \phn9.90 & \phn9.77 \\
4 & 12.38 & 11.72 & 12.84 & 10.25 & 10.00 \\
5 & 12.13 & 11.53 & 12.64 & 10.04 & \phn9.79 \\
6 & 11.91 & 11.39 & 12.52 & \phn9.88 & \phn9.64 \\
7 & 12.32 & 11.86 & 12.95 & 10.26 & 10.03 \\
8 & 12.10 & 11.66 & 12.74 & 10.04 & \phn9.81 \\
9 & 11.89 & 11.51 & 12.60 & \phn9.88 & \phn9.66 \\
10 & 12.35 & 11.65 & 12.72 & 10.23 & \phn9.98 \\
11 & 12.09 & 11.46 & 12.52 & 10.00 & \phn9.75 \\
12 & 11.87 & 11.33 & 12.40 & \phn9.86 & \phn9.62 \\
13 & 12.32 & 11.88 & 13.03 & 10.28 & 10.13 \\
14 & 12.30 & 11.85 & 12.95 & 10.25 & 10.03 \\
15 & 11.87 & 11.50 & 12.61 & \phn9.88 & \phn9.67 \\
16 & 12.29 & 11.85 & 12.96 & 10.27 & 10.04 \\
17 & 12.29 & 11.85 & 12.96 & 10.27 & 10.04 \\
18 & 12.10 & 11.54 & 12.61 & 10.03 & \phn9.74 \\
\enddata
\tablenotetext{a}{Note that the rows for models 16 and 17 are identical since
the same values for the parameters that affect cloud evaporation
($n_\mathrm{cl}$, $T_h$, $B_0$ and $N($\ion{H}{1}$)$), were used. The
difference between the models in this case is the different FUV radiation
fields used.  The difference in the FUV field has negligable effect on the
column densities of the high ions.}
\end{deluxetable}

\clearpage
%tab:abund
\begin{deluxetable}{crrrrrrr} 
\tablecolumns{8} 
\tablewidth{0pc} 
\tablecaption{Elemental Gas Phase Abundances (ppm)\label{tab:abund}} 
\tablehead{ 
\colhead{} & \multicolumn{7}{c}{Element} \\ 
\cline{2-8} \\
\colhead{Model No.} & \colhead{C} & \colhead{N} & \colhead{O} & 
\colhead{Mg} & \colhead{Si} & \colhead{S} & \colhead{Fe} \\ }
\startdata
 1 & 457 &  81.3 & 631 & 7.94 &  10.2 &  22.4 & 3.16 \\
 2 & 263 &  50.1 & 380 & 4.68 &  6.03 &  13.2 & 1.86 \\
 3 & 166 &  34.7 & 275 & 3.09 &  4.07 &  8.91 & 1.26 \\
 4 & 617 &  74.1 & 631 & 9.12 &  11.5 &  25.1 & 3.55 \\
 5 & 339 &  46.8 & 389 & 5.25 &  6.76 &  14.5 & 2.04 \\
 6 & 191 &  33.1 & 282 & 3.39 &  4.37 &  9.55 & 1.35 \\
 7 & 457 &  83.2 & 617 & 8.13 &  10.2 &  22.4 & 3.09 \\
 8 & 275 &  52.5 & 380 & 4.90 &  6.03 &  13.2 & 1.86 \\
 9 & 162 &  36.3 & 275 & 3.16 &  3.98 &  8.71 & 1.26 \\
10 & 617 &  75.9 & 631 & 9.33 &  11.5 &  25.1 & 3.47 \\
11 & 331 &  47.9 & 389 & 5.37 &  6.61 &  14.5 & 2.00 \\
12 & 191 &  33.9 & 282 & 3.47 &  4.37 &  9.33 & 1.35 \\
13 & 437 &  79.4 & 631 & 7.94 &  10.5 &  22.9 & 3.24 \\
14 & 437 &  81.3 & 617 & 8.13 &  10.2 &  22.9 & 3.16 \\
15 & 155 &  35.5 & 275 & 3.24 &  4.07 &  8.91 & 1.29 \\
16 & 427 &  81.3 & 631 & 8.32 &  10.5 &  23.4 & 3.24 \\
17 & 427 &  81.3 & 631 & 8.32 &  10.5 &  23.4 & 3.24 \\
18 & 309 &  49.0 & 389 & 5.37 &  6.61 &  14.5 & 2.04 \\
19 & 759 &  67.6 & 631 & 9.77 &  12.6 &  27.5 & 3.80 \\
20 & 398 &  43.7 & 389 & 5.62 &  7.08 &  15.5 & 2.14 \\
21 & 219 &  30.9 & 282 & 3.55 &  4.57 &  10.0 & 1.41 \\
22 & 724 &  72.4 & 631 & 10.0 &  12.0 &  26.9 & 3.63 \\
23 & 380 &  45.7 & 389 & 5.75 &  6.92 &  15.1 & 2.09 \\
24 & 209 &  32.4 & 275 & 3.63 &  4.47 &  9.77 & 1.38 \\
25 & 832 &  69.2 & 631 & 9.77 &  12.3 &  26.9 & 3.72 \\
\enddata
\end{deluxetable}

\clearpage
%tab:sunres
\begin{deluxetable}{cccccccccccc} 
\tabletypesize{\footnotesize}
\rotate
\tablecolumns{11}
\tablewidth{0pc} 
\tablecaption{Model Results for Solar Location\label{tab:sunres}}
\tablehead{ 
\colhead{Model} & \colhead{$X(\mathrm{H})$} & \colhead{$X(\mathrm{He})$} &
\colhead{\ion{O}{1}/\ion{N}{1}} & \colhead{\ion{O}{1}/\ion{He}{1}} &
\colhead{\ion{N}{1}/\ion{He}{1}} & \colhead{\ion{Ne}{1}/\ion{He}{1}} &
\colhead{\ion{Ar}{1}/Ar$_\mathrm{tot}$} & \colhead{$T$} &
\colhead{$n($\ion{H}{1}$)$} & \colhead{$n($\ion{He}{1}$)$} & \colhead{$n_e$}
\\}
\startdata
Obs.\tablenotemark{a} & \nodata & \nodata & 7.0\phn &
$3.7 \times10^{-3}$\phn & $5.2 \times10^{-4}$\phn & 
$6.0 \times10^{-4}$\phn & \nodata & $7000$ & \nodata & 0.017\phn & \nodata \\
\hline
1 & 0.309 & 0.467 & 9.41 & $8.51\times10^{-3}$ & $9.05\times10^{-4}$ &
$3.02\times10^{-4}$ & 0.180 & 7010 & 0.205 & 0.0156 & 0.106\phn \\
2 & 0.287 & 0.471 & 9.49 & $5.35\times10^{-3}$ & $5.64\times10^{-4}$ &
$2.86\times10^{-4}$ & 0.199 & 8230 & 0.208 & 0.0152 & 0.0982 \\
3 & 0.256 & 0.444 & 9.73 & $3.83\times10^{-3}$ & $3.93\times10^{-4}$ &
$2.98\times10^{-4}$ & 0.235 & 8640 & 0.233 & 0.0171 & 0.0949 \\
4 & 0.224 & 0.396 & 9.41 & $8.36\times10^{-3}$ & $8.88\times10^{-4}$ &
$3.45\times10^{-4}$ & 0.253 & 6150 & 0.235 & 0.0181 & 0.0806 \\
5 & 0.221 & 0.408 & 9.64 & $5.30\times10^{-3}$ & $5.50\times10^{-4}$ &
$3.18\times10^{-4}$ & 0.263 & 7750 & 0.225 & 0.0169 & 0.0757 \\
6 & 0.204 & 0.389 & 9.89 & $3.80\times10^{-3}$ & $3.85\times10^{-4}$ &
$3.18\times10^{-4}$ & 0.296 & 8400 & 0.239 & 0.0180 & 0.0735 \\
7 & 0.332 & 0.507 & 9.45 & $8.81\times10^{-3}$ & $9.32\times10^{-4}$ &
$2.48\times10^{-4}$ & 0.164 & 7380 & 0.195 & 0.0140 & 0.113\phn \\
8 & 0.300 & 0.511 & 9.42 & $5.76\times10^{-3}$ & $6.12\times10^{-4}$ &
$2.37\times10^{-4}$ & 0.182 & 8480 & 0.202 & 0.0137 & 0.101\phn \\
9 & 0.266 & 0.483 & 9.66 & $4.12\times10^{-3}$ & $4.27\times10^{-4}$ &
$2.47\times10^{-4}$ & 0.216 & 8870 & 0.228 & 0.0156 & 0.0993 \\
10 & 0.245 & 0.432 & 9.59 & $8.76\times10^{-3}$ & $9.13\times10^{-4}$ &
$2.80\times10^{-4}$ & 0.229 & 6650 & 0.223 & 0.0164 & 0.0868 \\
11 & 0.255 & 0.442 & 9.84 & $5.45\times10^{-3}$ & $5.54\times10^{-4}$ &
$2.63\times10^{-4}$ & 0.233 & 8080 & 0.212 & 0.0155 & 0.0861 \\
12 & 0.215 & 0.423 & 9.96 & $4.02\times10^{-3}$ & $4.03\times10^{-4}$ &
$2.64\times10^{-4}$ & 0.272 & 8640 & 0.235 & 0.0169 & 0.0782 \\
13 & 0.298 & 0.454 & 9.50 & $8.42\times10^{-3}$ & $8.87\times10^{-4}$ &
$3.13\times10^{-4}$ & 0.190 & 6920 & 0.229 & 0.0176 & 0.113\phn \\
14 & 0.319 & 0.493 & 9.51 & $8.70\times10^{-3}$ & $9.15\times10^{-4}$ &
$2.59\times10^{-4}$ & 0.174 & 7290 & 0.219 & 0.0159 & 0.120\phn \\
15 & 0.259 & 0.472 & 9.75 & $4.05\times10^{-3}$ & $4.16\times10^{-4}$ &
$2.58\times10^{-4}$ & 0.226 & 8810 & 0.252 & 0.0176 & 0.106\phn \\
16 & 0.312 & 0.484 & 9.63 & $8.82\times10^{-3}$ & $9.16\times10^{-4}$ &
$2.66\times10^{-4}$ & 0.180 & 7210 & 0.235 & 0.0173 & 0.125\phn \\
17 & 0.312 & 0.484 & 9.63 & $8.82\times10^{-3}$ & $9.16\times10^{-4}$ &
$2.66\times10^{-4}$ & 0.180 & 7210 & 0.235 & 0.0173 & 0.125\phn \\
18 & 0.234 & 0.448 & 9.56 & $5.65\times10^{-3}$ & $5.91\times10^{-4}$ &
$2.62\times10^{-4}$ & 0.240 & 8140 & 0.242 & 0.0170 & 0.0894 \\
19 & 0.180 & 0.313 & 9.60 & $7.74\times10^{-3}$ & $8.06\times10^{-4}$ &
$4.18\times10^{-4}$ & 0.322 & 4890 & 0.258 & 0.0214 & 0.0671 \\
20 & 0.183 & 0.346 & 9.78 & $5.01\times10^{-3}$ & $5.12\times10^{-4}$ &
$3.51\times10^{-4}$ & 0.321 & 7200 & 0.228 & 0.0180 & 0.0618 \\
21 & 0.186 & 0.319 & 10.2 & $3.47\times10^{-3}$ & $3.42\times10^{-4}$ &
$3.57\times10^{-4}$ & 0.350 & 8000 & 0.233 & 0.0193 & 0.0629 \\
22 & 0.207 & 0.385 & 9.55 & $8.45\times10^{-3}$ & $8.84\times10^{-4}$ &
$3.03\times10^{-4}$ & 0.274 & 6050 & 0.235 & 0.0179 & 0.0735 \\
23 & 0.226 & 0.395 & 9.85 & $5.20\times10^{-3}$ & $5.28\times10^{-4}$ &
$2.79\times10^{-4}$ & 0.274 & 7750 & 0.216 & 0.0165 & 0.0746 \\
24 & 0.211 & 0.371 & 9.83 & $3.60\times10^{-3}$ & $3.66\times10^{-4}$ &
$2.84\times10^{-4}$ & 0.307 & 8410 & 0.227 & 0.0177 & 0.0725 \\
25 & 0.194 & 0.341 & 9.55 & $7.96\times10^{-3}$ & $8.33\times10^{-4}$ &
$3.85\times10^{-4}$ & 0.295 & 5120 & 0.211 & 0.0170 & 0.0607 \\
\enddata
\tablenotetext{a}{Observational results from \citet{GloecklerGeiss:2001},
Gloeckler (2000) and \citet{Witte:1996} (see table \ref{tab:obs} for
uncertainties).}
\end{deluxetable}

\clearpage
%tab:ioniz
% RUN SUMMARY
% File name:intrfc+stars18.log
% Run Description:
% c  FUV/EUV/SXR backgrounds - in soft x-rays includes only enough        
% c  "additional" emission to account for B band rate                     
% c  data from /home/slavin/cldevap/SXRB_add.dat24                        
% c  cosmic abundances assumed in hot gas                                 
% c  Gondhalekar et al. (1980) FUV spectrum                               
% c  assumed HI column density =  4.00000e+17                             
% c  Emission from an evaporative boundary to the Local Cloud             
% c  data from /home/slavin/cldevap/cldspect2.dat24                       
% c  model has: Rcl = 3., Rf = 30., ncl = 0.35, Tf = 1.3E+06, B0 = 5.00,  
% c  Kred = 0.5, mdot = 0.669, XH(R) = 0.45, XHe(R) = 0.61                
% c  XH(0) = 0.30, XHe(0) = 0.53                                          
% c  LIC abundances -- with enhanced O, Mg and Fe (ref. Gry) and          
% c  additional changes for C, Si, Fe, and S                              
% c !New - abundances of C, N, O, Mg, Si and Fe set so that column densiti
% c will match the Gry & Dupin observations towards epsilon CMa (blue + LI
\begin{deluxetable}{llllll}
\tablecolumns{6}
\tablewidth{0pc}
\tablecaption{Model Predictions for Ionization Fractions at the Sun
\tablenotemark{a}\label{tab:ioniz}}
\tablehead{
\colhead{Element} & \colhead{Abundance} & \multicolumn{4}{c}{Ionization
Fraction} \\
\colhead{} & \colhead{(ppm)} & \colhead{I} & \colhead{II} 
& \colhead{III} & \colhead{IV}}
\startdata
H  & $\phn10^{6}$ &  0.688 &  0.312 & \nodata & \nodata \\
He & $\phn10^{5}$ &  0.506 &  0.484 & 0.0103 & \nodata \\
C  & 427 & 0.0005 &  0.964 & 0.0352 & 0.0 \\
N  & \phn81.3 & 0.570 &  0.430 & 0.0003 & 0.0 \\
O  & 631 & 0.707 & 0.293 & 0.0001 & 0.0 \\
Ne & 123 & 0.110 & 0.627 &  0.264 & 0.0 \\
Na & \phn\phn2.04 & 0.0028 &  0.865 &  0.132 & 0.0 \\
Mg & \phn\phn8.32 & 0.0020 &  0.827 &  0.171 & 0.0 \\
Al & \phn\phn0.0794 & 0.0001 &  0.973 & 0.0204 & 0.0066 \\
Si & \phn10.5 & 0.0 & 0.998 & 0.0022 & 0.0 \\
P  & \phn\phn0.219 & 0.0002 &  0.968 & 0.0314 & 0.0001 \\
S  & \phn23.4 & 0.0001 &  0.959 & 0.0413 & 0.0 \\
Ar & \phn\phn2.82 &  0.180 &  0.493 &  0.328 & 0.0 \\
Ca & \phn\phn0.000407 & 0.0 & 0.0247 &  0.975 & 0.0003 \\
Fe & \phn\phn3.24 & 0.0002 & 0.962 & 0.0378 & 0.0 \\
\enddata
\tablenotetext{a}{Results for model no.\ 17 as described in Table
\ref{tab:modparm}.}
\end{deluxetable}

\end{document}